\pdfoutput=1
\documentclass[arxiv,article,moreauthors,pdftex]{mdpi} 
\firstpage{1} 
\pubvolume{xx}
\issuenum{x}
\articlenumber{x}
\pubyear{2020}
\copyrightyear{2020}
\history{Received: date; Accepted: date; Published: date}

\usepackage{graphicx}
\usepackage[algoruled,vlined]{algorithm2e}
\usepackage{commath, amsmath}
\usepackage{xcolor}
\usepackage{hyperref}
\usepackage{natbib}

\DeclareMathOperator*{\argmin}{arg\,min}
\newcommand{\pfv}{d} 
\newcommand{\Gf}{G_\text{c}} 
\newcommand{\lc}{l_\text{c}} 
\newcommand{\Fcde}{\tilde{Y}}  
\newcommand{\ft}{\sigma_\text{c}} 
\newcommand{\Fref}[1]{Figure~\ref{#1}}
\newcommand{\Eref}[1]{Eq.~\eqref{#1}}

\newcommand{\Sref}[1]{Section~\ref{#1}}
\newcommand{\Tref}[1]{Table~\ref{#1}}

\newcommand{\de}[1]{\,{\mathrm d}#1} 
\newcommand{\der}[2]{\dod{#1}{#2}}
\newcommand{\ppd}[2]{\dpd{#1}{#2}}

\newcommand{\E}{E} 
\newcommand{\PR}{\nu} 

\Title{Phase-field Fracture Modelling of Thin Monolithic and Laminated Glass Plates under Quasi-static Bending}

\Author{Jaroslav Schmidt $^{1
}$\orcidA{}, 
Alena Zemanov\'{a} $^{1}$*\orcidB{}, 
Jan Zeman$^{1}$\orcidC{}, and 
Michal \v{S}ejnoha $^{1}$}

\AuthorNames{Jaroslav Schmidt, Alena Zemanov\'{a}, Jan Zeman, and Michal \v{S}ejnoha}

\address[1]{%
$^{1}$ \quad Department of Mechanics, Faculty of Civil Engineering, Czech Technical University in Prague,
Thakurova 7, 166 29 Prague 6 – Dejvice, Czech Republic}

\corres{Correspondence: alena.zemanova@fsv.cvut.cz
}

\abstract{
A phase-field description of brittle fracture is employed in the reported four-point bending analyses of monolithic and laminated glass plates. Our aims are: (i) to compare different phase-field fracture formulations applied to thin glass plates, (ii) to assess the consequences of the dimensional reduction of the problem and mesh density and refinement, and (iii) to validate for quasi-static loading the time/temperature-dependent material properties we derived recently for two commonly used polymer foils made of Polyvinyl Butyral or Ethylene-Vinyl Acetate. As the nonlinear response prior to fracture, typical of the widely used Bourdin-Francfort-Marigo model, can lead to a significant overestimation of the response of thin plates under bending, the numerical study investigates two additional phase-field fracture models providing the linear elastic phase of the stress-strain diagram. The typical values of the critical fracture energy and tensile strength of glass lead to a phase-field length-scale parameter that is challenging to resolve in the numerical simulations. So, we show how to determine the fracture energy concerning the applied dimensional reduction and the value of the length-scale parameter relative to the thickness of the plate. The comparison shows that the phase-field models provide very good agreement with the measured stresses and resistance of laminated glass, despite the fact that only one/two cracks localised using the quasi-static analysis, whereas multiple cracks evolved during the experiment. It has also been observed that the stiffness and resistance of the partially fractured laminated glass can be well approximated using a 2D plane-stress model with initially predefined cracks, which provides a better estimation than the one-glass-layer limit.}

\keyword{
laminated glass; phase-field fracture model; quasi-static bending; 
PVB; EVA
}

\begin{document}

\section{Introduction}
\label{S:INT}

 Glass, despite its brittleness, has proved to be a suitable material for load-bearing or fail-safe transparent structures when combined with polymers or other plastic interlayers~\cite{Haldimann:2008:SUG, ledbetter2006structural}. Laminated glass dominates the structural applications of glass as it preserves the optical qualities of glass, and above that, it provides a certain level of post-breakage resistance~\cite{zhao2019experimental}, e.g., after an impact, an earthquake, or another disaster. Especially in the case of an unforeseen event, lamination can increase structural reliability significantly~\cite{bonati2019redundancy}.

%
\begin{figure}[ht]
    \centering
    \begin{tabular}{c}
        \includegraphics[width=0.8\textwidth]{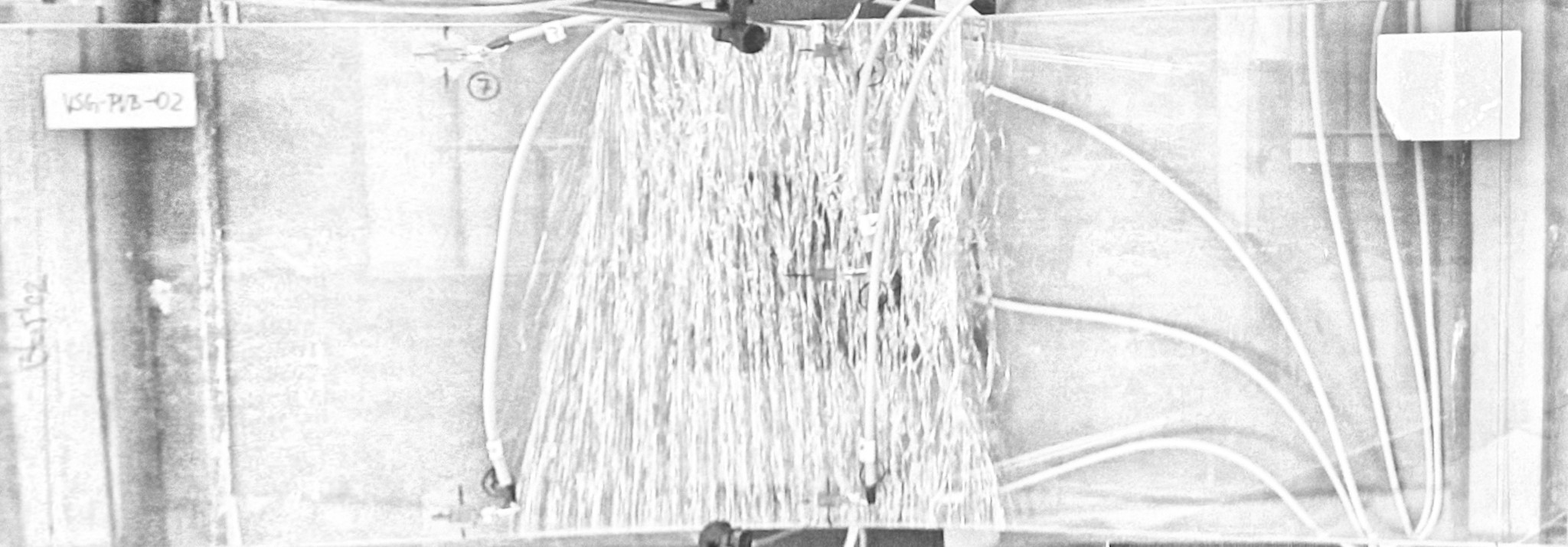} 
        \\ \small{four-point bending} 
        \\
        \includegraphics[width=0.8\textwidth]{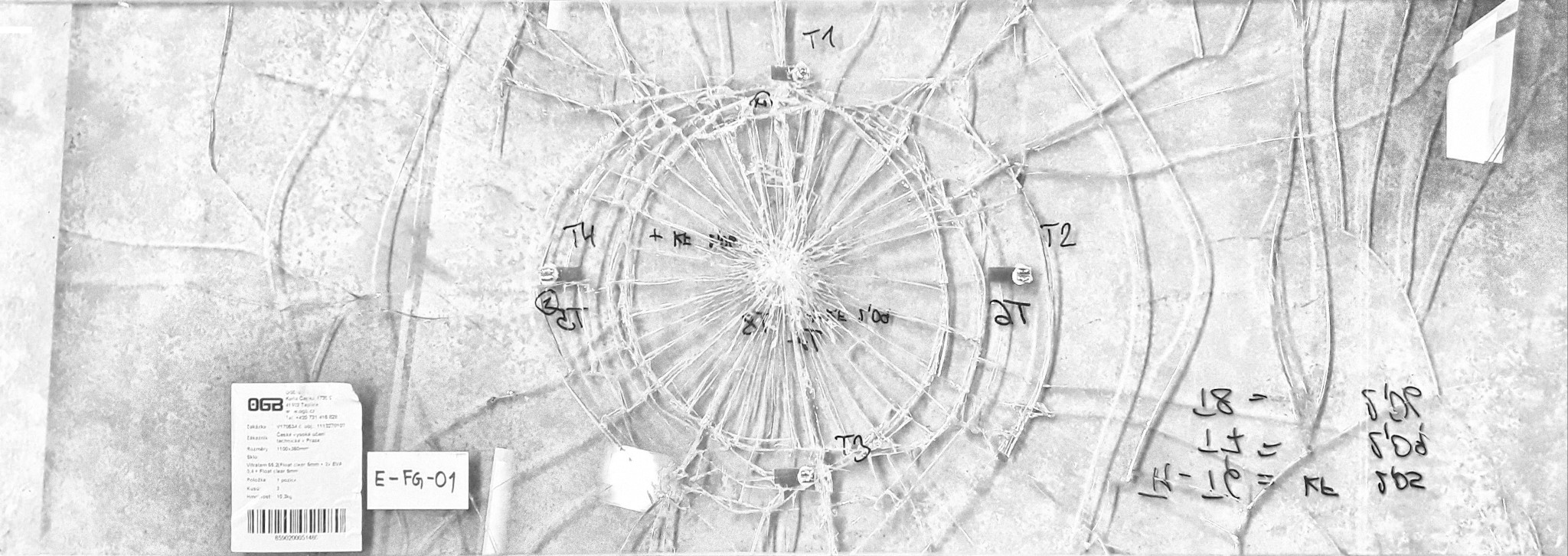}
        \\ \small{low-velocity hard impact} 
    \end{tabular}
    \caption{Photos of fracture patterns for laminated glass samples made of float glass with two-edge supports after four-point bending with multiple cracks in the middle part or after low-velocity hard impact of a steel impactor with a hemispherical head resulting in a spiderweb crack pattern.}
    \label{fig:PFP}
\end{figure}
To identify the points of crack initiation and predict their propagation and subsequent response of glass elements is 
of paramount importance
for understanding the structural performance and design of laminated glass plates. The resulting fracture pattern indicates the cause of a glass failure and therefore provides valuable information for its diagnostic interpretation,~\Fref{fig:PFP}. Different crack patterns are typical of glass plates broken due to over-stressing caused by a uniform loading, thermal stresses, a hard or soft impact or due to a nickel inclusion or an instability failure~\cite{overend2007diagnostic}.

In the literature, recent research papers in this field illustrate the effort of numerous researchers to simulate and predict the pre-fracture, fracture, and post-fracture behaviour of laminated glass~\cite{teotia2018applications, chen2017numerical}. 
For the practical design, effective thickness approaches were derived from an analytical model to predict extreme values of deflections or tensile stresses~\cite{calderone2009effective, galuppi2012effective, galuppi2012effectiveP, galuppi2014enhanced}. 
Except for the methods replacing the multi-layer glass element with an equivalent monolithic one, analytical models for two plates under bending with a shear coupling provided by the 
interlayer were derived for laminated glass beams and plates~\cite{vallabhan1993analysis, acsik2003laminated, acsik2005mathematical,ivanov2006analysis,foraboschi2007behavior,koutsawa2007static,foraboschi2012analytical}. Analytical models based on layer-wise type theories were also tested for photovoltaic panels~\cite{schulze2012analysis}. 
For numerical modelling, a comprehensive review of the finite element method in failure analysis of laminated glass can be found in~\cite{teotia2018applications}.

%
Predicting and modelling the crack initiation and propagation in materials and structures has been one of the most significant challenges in solid mechanics for decades. In recent years, modelling based on the phase-field fracture/damage formulations has become an elegant, powerful, and well-established tool to simulate the fracture in brittle solids. 
The mathematical structure of phase-field formulations of fracture is similar to the continuum gradient-damage models, but their components are interpreted differently. 
The phase-field methods can predict crack initiation and deal with intricate crack patterns like branching. Within the scope of the widely used finite element framework, the phase-field models provide promising results for many engineering problems, e.g.,~\cite{lancioni2009variational, martinez2018phase, natarajan2019modeling}, and became popular in the computational modelling of fracture
including their recent application to tall laminated glass beams under in-plane tension and bending~\cite{freddi2020phase}. 
On the other hand, the phase-field models, despite their straightforward and relatively simple implementation, are computationally demanding due to the introduction of new variables in nodes corresponding to the phase-field parameters and due to fine meshes needed. 

In this paper, we focus on the pre- and post-fracture bending of laminated glass. 
We aim at predicting the flexural response of multi-layer samples in terms of, e.g., deflections and stresses, and discuss the corresponding fracture pattern predicted by the numerical model.
This way, we want to assess the strengths and drawbacks of the phase-field formulation in fracture modelling of laminated glass under quasi-static bending and their usability for practical structural elements. Therefore, we demonstrate the performance of selected formulations on examples of thin rectangular plates of dimensions 1,100 $\times$ 360~mm$^2$ under bending, made of monolithic or laminated glass with two opposite sides simply-supported. 
The presented phase-field approaches differ in the sharp crack approximation and the employed crack driving forces.
Due to the fine mesh needed for the phase-field fracture analysis,
different dimensional-reduction strategies for the finite element discretization were utilized, i.e., multi-layer beam or plate models and a 2D sectional plane-stress model. For the interlayer, we tried to replace the originally viscoelastic time/temperature-dependent model~\cite{hana2019experimental} with an elastic one
to simplify the simulations, while still obtaining reliable results.

The paper is structured as follows. Phase-field formulations tested for glass fracture are introduced in~\Sref{S:PFF}. The parameters for the numerical models are discussed, and their relationships are derived in~\Sref{S:Par}. Then, the formulations are compared for a monolithic glass plate with considering differently discretizations in~\Sref{S:NCS}. The experimental program on laminated glass in bending is reviewed in~\Sref{S:ETS}, and the numerical predictions are validated in~\Sref{S:ESV}. Finally, the main conclusions from this study are summarized in~\Sref{S:CON}. 

\section{Phase-field formulation for glass fracture} 
\label{S:PFF}

\subsection{Overview of selected approaches}

Phase-field damage formulations can be seen as a generalisation of Griffith's theory of brittle fracture~\cite{griffith1921vi}. The formulations are based on the variational approximation~\cite{bourdin2000numerical} of brittle fracture proposed in~\cite{francfort1998revisiting}. The aim of the phase-field models is to predict the displacement field and the crack evolution by the minimisation of the total potential energy, and therefore, no other assumption for cracks initiation is needed. For this purpose, a minimal set of input parameters is required, i.e., the experimentally measurable Young modulus, Poisson ratio, and critical fracture energy, plus a length-scale parameter, which can be seen as a pure numerical parameter for sharp crack regularization or as a material parameter determined under suitable assumptions from the tensile strength of the brittle material~\cite{amor2009regularized}.  
%
%
A comprehensive summary focused on the phase-field modelling of fracture can be found in~\cite{wu2018phase}.
%
The individual phase-field models proposed in the literature differ in the way how the sharp crack is approximated, which degradation function is used to lower the stiffness with an increasing phase-field parameter,
or how the irreversibility condition is enforced to avoid unphysical self-healing. These models, originally describing brittle fracture, were extended to a cohesive and ductile fracture, or a dynamic response. Formulation are available for one-, two-, or three-dimensional problems, for structural elements including plates and shells~\cite{kiendl2016phase}, or for multi-physics problems~\cite{miehe2015phaseII}. 
In this study, we compare three phase-field formulations for glass fracture: 
\begin{itemize}
\itemsep0em 
    \item classical \textit{Bourdin-Francfort-Marigo's formulation with energetic criterion} (PF-B)~\cite{bourdin2000numerical, bourdin2008variational} extended to an anisotropic model with a tension/compression asymmetry by Amor et al.~\cite{amor2009regularized} or Miehe et al.~\cite{miehe2010thermodynamically}, 
    \item modified \textit{Miehe-Sch\"anzel-Ulmer's formulation with stress-based Rankine-type criterion} (PF-M)~\cite{miehe2015phase}, 
    \item \textit{Pham-Amor-Marigo-Maurini's energy-based formulation} for damage evolution (PF-P)~\cite{pham2011gradient}.
\end{itemize}
The PF-B formulation was selected for two reasons: First, it represents a widely used phase-field formulation for brittle fracture applied to engineering problems as the admissible range for the phase-field variable is intrinsically guaranteed for a quadratic degradation function. Second, to illustrate the drawback of this method in predicting the damage evolution right at the onset of loading, so the stress-strain diagram lacks an elastic phase~\cite{mandal2019length}. 

The other two formulations, i.e., PF-M and PF-P, try to overcome the drawback of the missing elastic domain. The PF-M model replaces for this purpose the original energy-based criterion with a principal tensile stress criterion with a threshold. This formulation with a Rankine-based crack driving force becomes variationally inconsistent, but the elastic properties in uncracked zones are preserved.
On the other hand, the PF-P formulation relies on the combination of a linear function 
for the approximation of the dissipated energy with a quadratic function for the degradation of the initial elastic energy. Then, the damage distribution in the diffuse localisation band has a parabolic shape, and the global response is linear in the initial stage. In contrast with the previous models, the admissible range for the phase-field parameter is not intrinsically satisfied, and the lower bound of this range has to be enforced by additional strategies. 

Undoubtedly, many other formulations can be found in the literature. For example, alternative degradation functions which do not yield  a pronounced softening response before the fracture initiation were systematically studied in~\cite{kuhn2015degradation}.

\subsection{Energy funcional for elastic bodies with sharp cracks}

In phase-field models, the displacements $\boldsymbol{u}$ together with the cracks, represented by surfaces in the domain of the system $\Gamma \subset \Omega$, are governed by the minimisation of the energy functional with respect to all admissible displacement fields $\hat{\boldsymbol{u}}$ and crack surfaces $\hat{\Gamma}$, i.e.,
\begin{eqnarray}
(\boldsymbol{u},\Gamma)=\argmin_{(\hat{\boldsymbol{u}},\hat{\Gamma})
}\{\mathcal{E}(\hat{\boldsymbol{u}},\hat{\Gamma})\}.
\end{eqnarray}
According to Griffith's theory~\cite{griffith1921vi}, this energy functional consists of three parts
\begin{equation}
\mathcal{E}({\boldsymbol u},\Gamma)=
\Psi_\text{e}({\boldsymbol u}, {\Gamma}) + \Psi_\text{s}(\Gamma)
-\mathcal{P}({\boldsymbol u}).
\label{E:EF}
\end{equation}
The first term stores the elastic strain energy,
\begin{equation}
    \Psi_\text{e}({\boldsymbol u}, {\Gamma}) = \int_{\Omega\setminus\Gamma}{\psi_\text{e}({\boldsymbol\varepsilon({\boldsymbol u})})\de{V}},
\end{equation}
where $\psi_\text{e}$ represents the elastic energy density with the strains defined as ${\boldsymbol\varepsilon} = \nabla_\text{sym}\boldsymbol{u}$. The second term is the dissipated surface energy,
\begin{equation}
    \Psi_\text{s}(\Gamma) = \Gf\int_\Gamma{\de A},
\end{equation}
where $\Gf$ is the Griffith-type fracture energy-per-unit-area or critical energy release rate of the material, in what follows fracture energy for short. The last term is the external potential energy functional $\mathcal{P}$ given by
\begin{equation}
    \mathcal{P}({\boldsymbol u})= \int_{\Omega}{\boldsymbol b^*} \cdot {\boldsymbol u} \de {V} + \int_{\partial\Omega_t}  {\boldsymbol t^*} \cdot {\boldsymbol u}\de A,
\end{equation}
where $\boldsymbol b^*$ are the distributed body forces in $\Omega$ and $\boldsymbol t^*$ are the tractions applied on the external boundary $\partial\Omega_t$.

\subsection{Phase-field approximation}
The next step is an approximation of the sharp discontinuity with a diffuse crack.
Then, the dissipated energy $\Psi_\text{s}(\Gamma)$ can be approximated by a volume integral
\begin{equation}
\Gf \int_{\Gamma}\de A\approx\int_\Omega\psi_{\text{s}}(\pfv,\nabla \pfv)\de V
= 
\frac{\Gf}{c_{\alpha}} \int_\Omega \frac{1}{\lc}\alpha(\pfv) + \lc | \nabla \pfv |^2\de V.
\end{equation}
The phase-field variable $\pfv$ characterises the state of the material, so that $\pfv = 0$ corresponds to an intact material and $\pfv = 1$ to a
fully cracked material. The geometric function $\alpha(\pfv)$ characterises the evolution of the damage $\pfv$, and the scaling parameter $c_{\alpha}$ is given by $c_{\alpha} = 4 \int_0^1 \sqrt{\alpha(\beta)}\de\beta$, see ahead to~\Tref{tab:par_mod} on Page~\pageref{tab:par_mod}. The length-scale parameter
$\lc$ controls the width of the diffuse interface, i.e., of the continuous approximation to the discrete crack.

\begin{figure}[ht]
    \centering
    \begin{tabular}{cc}
    \small
    \def\svgwidth{0.45 \textwidth}
\begingroup%
  \makeatletter%
  \providecommand\color[2][]{%
    \errmessage{(Inkscape) Color is used for the text in Inkscape, but the package 'color.sty' is not loaded}%
    \renewcommand\color[2][]{}%
  }%
  \providecommand\transparent[1]{%
    \errmessage{(Inkscape) Transparency is used (non-zero) for the text in Inkscape, but the package 'transparent.sty' is not loaded}%
    \renewcommand\transparent[1]{}%
  }%
  \providecommand\rotatebox[2]{#2}%
  \newcommand*\fsize{\dimexpr\f@size pt\relax}%
  \newcommand*\lineheight[1]{\fontsize{\fsize}{#1\fsize}\selectfont}%
  \ifx\svgwidth\undefined%
    \setlength{\unitlength}{226.77165354bp}%
    \ifx\svgscale\undefined%
      \relax%
    \else%
      \setlength{\unitlength}{\unitlength * \real{\svgscale}}%
    \fi%
  \else%
    \setlength{\unitlength}{\svgwidth}%
  \fi%
  \global\let\svgwidth\undefined%
  \global\let\svgscale\undefined%
  \makeatother%
  \begin{picture}(1,0.3125)%
    \lineheight{1}%
    \setlength\tabcolsep{0pt}%
    \put(0,0){\includegraphics[width=\unitlength,page=1]{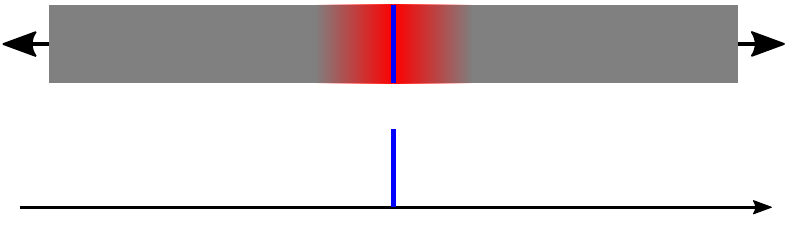}}%
    \put(0.49954782,0.15644112){\color[rgb]{0,0,0}\makebox(0,0)[lt]{\lineheight{1.25}\smash{\begin{tabular}[t]{l}1\end{tabular}}}}%
    \put(0,0){\includegraphics[width=\unitlength,page=2]{figure02a.pdf}}%
    \put(0.50002584,0.00427895){\color[rgb]{0,0,0}\makebox(0,0)[t]{\lineheight{1.25}\smash{\begin{tabular}[t]{c}0\end{tabular}}}}%
  \end{picture}%
\endgroup%

         &
    \includegraphics{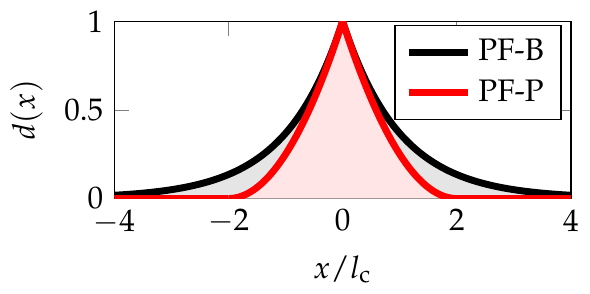}
    \end{tabular}
    \caption{Schematic diffuse crack approximation of originally sharp discontinuity and the detail of the damage profiles for two different geometric crack functions corresponding to  a one-dimensional localised solution of an infinite bar under tension, i.e., $d=\exp(-|x|/\lc)$ for the PF-B model or $d=(1-0.5|x|/\lc)^2$ for $|x| \leq 2\lc$ and $d=0$ otherwise for the PF-P model.}
    \label{fig:dam_prof}
\end{figure}

Examples of the one-dimensional damage profile for variationally consistent models, PF-B and PF-P, are plotted in~\Fref{fig:dam_prof}. If the length scale $\lc$ is seen as a material parameter, its relationships with fracture energy $\Gf$ and tensile strength $\ft$ can be derived from a spatially homogeneous analytical solution, see~\cite{zhang2017numerical}. For beams and plates under bending, we will discuss its setting later in~\Sref{S:Par}.

Due to the crack regularization, the stored strain energy functional can be extended to the whole domain $\Omega$ using a degradation function $g(\pfv)$. Thus, the functional depends on both the displacements and phase-field variable
\begin{equation}
    \Psi_\text{e}({\boldsymbol u}, \pfv) =  \int_{\Omega} {g(\pfv)\psi_\text{e}({\boldsymbol\varepsilon({\boldsymbol u})})}\de V,
    \label{eq:anis}
\end{equation}
and the corresponding energy functional reads as
\begin{eqnarray}
\mathcal{E}({\boldsymbol u},\pfv) &=&  \int_{\Omega} {g(\pfv)\psi_\text{e}({\boldsymbol\varepsilon({\boldsymbol u})})
    }\de V \nonumber \\
    &+&
\frac{\Gf}{c_{\alpha}}\int_{\Omega}\frac{1}{\lc}\alpha(\pfv) + \lc | \nabla \pfv |^2\de V 
-
\int_{\Omega}{\boldsymbol b^*} \cdot {\boldsymbol u} \de V
-
\int_{\partial\Omega_t} {\boldsymbol t^*} \cdot {\boldsymbol u}\de A.
\label{eq:EF_withoutsplit}
\end{eqnarray}

\subsection{Anisotropic formulation}
\label{S:anis}
Furthermore, the elastic energy needs to be decomposed into the tensile and compressive contributions, to assure that the material cracks only under tension. 
Two conventional approaches available in the literature are the {volumetric-deviatoric split} and the {spectral decomposition},~\Tref{T:split}. 
The \textit{volumetric-deviatoric split} is based on the degradation of the energy incorporating the positive part of the volumetric strains and the total deviatoric strains~\cite{amor2009regularized}, whereas in the \textit{spectral decomposition}, the crack evolution is induced by the positive principal strains~\cite{miehe2010thermodynamically}.
\begin{table}[ht]
    \centering
    \begin{tabular}{ccc}
    \hline
        Split & $\psi_\text{e}^+({\boldsymbol\varepsilon})$ & $\psi_\text{e}^-({\boldsymbol\varepsilon})$
        \\
        \hline
        VD & $K/2\langle\text{tr}({\boldsymbol\varepsilon})\rangle^2 + \mu{\boldsymbol\varepsilon}_\text{D}:{\boldsymbol\varepsilon}_\text{D}$ & $K/2\langle-\text{tr}({\boldsymbol\varepsilon})\rangle^2$ \\
        SD & $\lambda/2\langle\text{tr}({\boldsymbol\varepsilon})\rangle^2 + \mu{\boldsymbol\varepsilon^+}:{\boldsymbol\varepsilon^+}$ & $\lambda/2\langle-\text{tr}({\boldsymbol\varepsilon})\rangle^2 + \mu{\boldsymbol\varepsilon^-}:{\boldsymbol\varepsilon^-}$ \\
        \hline
    \end{tabular}
        \caption{Strain energy decompositions: volumetric-deviatoric split (VD) and spectral decomposition (SD), where $K$ stands for the bulk modulus and $\lambda, \mu$ for Lame's coefficients, ${\boldsymbol\varepsilon}_\text{D}$ is the deviatoric part of the strain tensor, the Macaulay brackets are defined as $\left\langle a \right\rangle = \left( a + |a| \right)/2$, and the positive/negative components ${\boldsymbol\varepsilon^\pm}=\sum_i\pm\langle\pm\varepsilon_i\rangle\mathbf{p}_i\otimes\mathbf{p}_i$ follow from the principal strains $\varepsilon_i$ and the dyadic products of the eigenvectors $\mathbf{p}_i$ of ${\boldsymbol\varepsilon}$. }
    \label{T:split}
\end{table}

Consequently, only the tensile part of the strain energy functional is degraded, yielding
\begin{equation}
    \Psi_\text{e}({\boldsymbol u}, \pfv) =  \int_{\Omega} {g(\pfv)\psi^+_\text{e}({\boldsymbol\varepsilon({\boldsymbol u})})
    + \psi^-_\text{e}({\boldsymbol\varepsilon({\boldsymbol u})})
    }\de V.
    \label{E:split}
\end{equation}
Altogether, the anisotropic regularised energy function takes the form
\begin{eqnarray}
\mathcal{E}({\boldsymbol u},\pfv) &=&  \int_{\Omega} {g(\pfv)\psi^+_\text{e}({\boldsymbol\varepsilon({\boldsymbol u})})
    + \psi^-_\text{e}({\boldsymbol\varepsilon({\boldsymbol u})})
    }\de V \nonumber \\
    &+&
\frac{\Gf}{c_{\alpha}}\int_{\Omega}\frac{1}{\lc}\alpha(\pfv) + \lc | \nabla \pfv |^2\de V 
-
\int_{\Omega}{\boldsymbol b^*} \cdot {\boldsymbol u} \de V
-
\int_{\partial\Omega_t} {\boldsymbol t^*} \cdot {\boldsymbol u}\de A,
\label{eq:EF}
\end{eqnarray}
compare with~\Eref{eq:EF_withoutsplit}. Note that the anisotropy refers here to the formulation with the tension/compression split and not to an anisotropic nature of the material.

\subsection{Governing equations}

The unknown displacements and phase-field damage parameters are found by minimising the energy functional in~\Eref{eq:EF} complemented with boundary conditions. 
Thus, the governing equations describing the displacement sub-problem follow from~\Eref{eq:EF} by taking variation with respect to $\boldsymbol u$,
\begin{eqnarray}
    \nabla \cdot \boldsymbol{\sigma} + {\boldsymbol b^*} =\boldsymbol{0} && \text{in }\Omega,
    \nonumber \\
    \boldsymbol{\sigma} \cdot \boldsymbol{n} = {\boldsymbol t^*} && \text{on }\partial\Omega_t,
    \label{E:eq_eq}
\end{eqnarray}
where $\boldsymbol{n}$ is the outward unit normal vector to the boundary $\partial\Omega$. The stress field is given by
\begin{eqnarray}
 \boldsymbol{\sigma} = g(\pfv)\ppd{\psi^+_\text{e}({\boldsymbol\varepsilon({\boldsymbol u})})}{\boldsymbol\varepsilon} + \ppd{\psi^-_\text{e}({\boldsymbol\varepsilon({\boldsymbol u})})}{\boldsymbol\varepsilon}.
 \label{eq:stress}
\end{eqnarray}
The phase-field sub-problem yields the damage evolution equation and the corresponding Neumann boundary condition 
\begin{eqnarray} 
\left.
\begin{array}{ccc}
\displaystyle\frac{1}{c_{\alpha}}
\left(
\der{\alpha(\pfv)}{\pfv}
-2 \lc^2 \Delta \pfv \right) = 
- \frac{1}{2}
\der{g(\pfv)}{\pfv}
\Fcde
&&
\dot{\pfv} 
> 0
\\
\displaystyle\frac{1}{c_{\alpha}}
\left(
\der{\alpha(\pfv)}{\pfv}
-2 \lc^2 \Delta \pfv \right) > 
- \frac{1}{2}
\der{g(\pfv)}{\pfv}
\Fcde
&&
\dot{\pfv}
= 0
\end{array}
\right\}
&& \text{in }\Omega, \label{eq:PFGE}
\nonumber
\\
\left.
\begin{array}{ccc}
\frac{\Gf}{c_{\alpha}}2 \lc \nabla \pfv \cdot \boldsymbol{n} = {0}
&&
\dot{\pfv} 
> 0
\\
\frac{\Gf}{c_{\alpha}}2 \lc \nabla \pfv \cdot \boldsymbol{n} > {0}
&& 
\dot{\pfv} 
= 0
\end{array}
\right\}
&& \text{on }\partial\Omega,
\label{E:ev_eq}
\end{eqnarray}
where 
$\Delta$ is the Laplace operator,
$\dot{\pfv}$ denotes the damage rate, i.e. the derivative of the phase-field variable with respect to a (pseudo)time and
$\Fcde$ is a normalized effective damage/crack driving force.

\subsection{Specification of selected approaches}

Different damage models can be derived from this general formulation by a different choice of the geometric function $\alpha$ and the degradation function $g$, or by modification of the normalized effective crack driving force $\Fcde$, as shown in \Tref{tab:par_mod}.

\begin{table}[ht]
    \centering
    \begin{tabular}{cccccc}
    \hline
        Model & $\alpha(\pfv)$ & $c_{\alpha}$ &  $g(\pfv)$ & criterion & $\Fcde$
        \\ 
        \hline
        PF-B & $\pfv^2$ & 2 & $(1-\pfv)^2$ 
        & energetic 
        & $\frac{2\psi^+_\text{e}({\boldsymbol\varepsilon({\boldsymbol u})})}{\Gf/\lc}$
        \\
        PF-M & $\pfv^2$ & 2 & $(1-\pfv)^2$ 
        & Rankine-based -- stress
        & $\left\langle\frac{1}{\sigma_\text{c}^2} \sum_{i} \left\langle\bar{\sigma}_i\right\rangle^2-1 \right\rangle$
        \\
        PF-P & $\pfv$ 
        & ${8}/{3}$ & $(1-\pfv)^2$ 
        & energetic  
        &
        $\frac{2\psi^+_\text{e}({\boldsymbol\varepsilon({\boldsymbol u})})}{\Gf/\lc}$
        \\
        \hline
    \end{tabular}
        \caption{Phase-field formulations and corresponding functions and parameters. For the PF-M model, the Macaulay brackets are defined as $\left\langle a \right\rangle = \left( a + |a| \right)/2$, $\sigma_\text{c}$ is a critical fracture stress corresponding to the tensile strength, and $\bar{\sigma}_i$ are the principal components of the effective stress tensor.
        }
    \label{tab:par_mod}
\end{table}

In the PF-M formulation, the normalized effective driving force $\Fcde$ in~\Eref{E:ev_eq} is replaced by a Rankine-based one in terms of the principal effective stresses, \Tref{tab:par_mod} and \cite{miehe2015phase}, with the effective stress tensor given by
\begin{eqnarray}
 \boldsymbol{\bar{\sigma}} = \ppd{\psi_\text{e}({\boldsymbol\varepsilon({\boldsymbol u})})}{\boldsymbol\varepsilon}
 =
 \ppd{\psi^+_\text{e}({\boldsymbol\varepsilon({\boldsymbol u})})}{\boldsymbol\varepsilon} + \ppd{\psi^-_\text{e}({\boldsymbol\varepsilon({\boldsymbol u})})}{\boldsymbol\varepsilon},
\end{eqnarray}
compare with the stress tensor in~\Eref{eq:stress}.
The PF-M model is variationally inconsistent, but it can capture the asymmetric cracking behaviour with the initial linear elastic phase.

\subsection{Staggered scheme and hybrid formulation}
\label{S:hybrid}

The system of Eqs.~\eqref{E:eq_eq} and \eqref{E:ev_eq} can be solved using a \textit{staggered solution scheme}
based on  the alternating minimisation method, e.g.,~\cite{bourdin2000numerical,farrell2017linear}. 
For this iterative scheme, the nodal displacements are solved first using the crack phase-field damage fixed from the previous iterative step. Next, the updated displacements are utilized to solve for the phase-field variable, see ahead to Algorithm~\ref{alg:staggered} on page~\pageref{alg:staggered}.

Due to the tension/compression split of the strain energy discussed in~\Sref{S:anis}, the introduced \textit{anisotropic formulation} leads to nonlinear equilibrium equations~\eqref{E:eq_eq} to be solved numerically. To overcome this drawback and reduce the computational cost, a \textit{hybrid (isotropic-anisotropic) model} was introduced in~\cite{ambati2015review}. In this approach, 
the stress in~\Eref{eq:stress} is replaced by a term
\begin{eqnarray}
 \boldsymbol{\sigma} = g(\pfv)\ppd{\psi_\text{e}({\boldsymbol\varepsilon({\boldsymbol u})})}{\boldsymbol\varepsilon},
 \label{eq:stress_l}
\end{eqnarray}
that depends linearly on $\boldsymbol u$, cf.~\Eref{eq:EF_withoutsplit}. However, the phase-field evolution
is still driven by the normalized driving forces $\Fcde$ in \Eref{E:ev_eq}, to avoid cracks in zones under compression. 
See ahead on page~\pageref{alg:staggered_hyb} and compare Algorithms~\ref{alg:staggered} and \ref{alg:staggered_hyb}.

To conclude this section, let us highlight our assumption that the phase-field model for fracture is applied for glass layers only, whereas no damage initiates and evolves in the interlayer foils. Also no delamination on the glass/polymer interfaces is allowed, so we focus purely on the glass fracture in the multi-layer laminated glass plates.  

\section{Parameters for the phase-field models and their relation} 
\label{S:Par}

For the numerical analysis, we present in this section material parameters needed to predict the crack initiation and propagation in glass layers.
The elastic properties and tensile strength of glass in~\Tref{T:InputG} were set according to standards~\cite{en2004572}. The phase-field fracture formulation contains two additional parameters which have to be specified, i.e., the fracture energy $\Gf$ and the length-scale parameter $\lc$~\Eref{eq:EF}. 

\begin{table}[ht]
\centerline{
\begin{tabular}{l c r l}
\hline
Glass\\
\hline
Young's modulus of elasticity & $\E$ & 70 & GPa\\
Poisson's ratio & $\PR$ & 0.22 & --\\
Tensile strength & $\ft$ & 45 & MPa \\
\hline
\end{tabular}
}
\caption{Elastic material parameters and tensile strength of glass~\cite{en2004572}}
\label{T:InputG}
\end{table}

A typical value of the fracture energy  $\Gf=4$~J$\cdot$m$^{-2}$ for soda-lime-silica glass can be found in the literature~\cite{wiederhorn1969fracture,wang2017comparative}. Then, the corresponding length-scale parameter $\lc$ seen as a material parameter can be estimated. For illustration, we adopt the following analytical expressions, see~\Tref{T:lc}, 
\begin{eqnarray}
\lc &= \displaystyle\frac{27}{256}\displaystyle\frac{E \Gf}{\ft^2} &\hspace{5mm} \text{for PF-B model~\cite{bourdin2008variational, pham2017experimental,zhang2017numerical}}, 
\label{eq:lcB}
\\
\lc &= \displaystyle\frac{3}{8}\displaystyle\frac{E \Gf}{\ft^2} &\hspace{5mm} \text{for PF-P model~\cite{wu2018length,wu2018phase}},   
\label{eq:lcP}
\end{eqnarray}
derived from spatially homogeneous solutions of the one-dimensional quasi-static problems. This results in a very small length-scale parameter of about 10~$\mu$m, and therefore, leads to a very fine mesh and excessive computational costs for our real-size samples or glass panels in structural glass facades. 
For this reason, we decided to treat the length scale as a numerical parameter and to derive the corresponding fracture energy from Eqs.~\eqref{eq:lcB} and~\eqref{eq:lcP}.
For example, the fracture energies corresponding to the length-scale parameter of 3~mm can be seen in~\Tref{T:lc}. This way, the length-scale parameter can be set according to the mesh density, and the fracture energy can be adjusted for the simulation.     


\begin{table}[ht]
\centering
{
\begin{tabular}{cccccc}
\hline
 & 
 $\Gf$~\cite{wang2017comparative} & $\lc = \frac{27}{256}\frac{E \Gf}{\ft^2}$ & $\lc = \frac{3}{8}\frac{E \Gf}{\ft^2}$ & $\Gf = \frac{256}{27}\frac{f_\text{t}^2 \lc}{E}$ & $\Gf = \frac{8}{3}\frac{\ft^2 \lc}{E}$\\
 &
 [J$\cdot$m$^{-2}$] &  [mm] &  [mm] & [J$\cdot$m$^{-2}$] & [J$\cdot$m$^{-2}$]\\
\hline
material parameter & 4 & 0.010 & 0.036\\
numerical parameter &   & 3 & 3 & 823 & 231 \\
\hline
\end{tabular}
}
\caption{Link between the fracture energy $\Gf$ and the length-scale parameter $\lc$ seen as material or numerical parameter using the Young modulus $E=70$~GPa and the tensile strength for annealed glass $\ft=45$~MPa from~\cite{en2004572}} 
\label{T:lc}
\end{table}

Although derived for pure tension, the relationships introduced in~\Tref{T:lc} could also be used for thin structures in bending providing the stress distribution is close to constant within the finite elements in which the glass fracture could occur. Therefore, 
this approximation was used in the 2D~plane-stress model of the longitudinal cross-section where the mesh was refined in the area of maximal bending moments.

On the contrary, the introduced relations in Eqs.~\eqref{eq:lcB} and \eqref{eq:lcP} are not applicable if we employ beam and plates formulations with one element per the layer thickness characterized by a through-the-thickness constant damage parameter. 
To illustrate this fact, we establish relations between the length-scale parameter $\lc$ and the fracture energy $\Gf$ for beams under pure bending. 

Similarly to previous approaches, e.g.~\cite{borden2012phase, pham2017experimental, zhang2017numerical}, 
we start from the evolution equation for the damage distribution~\Eref{E:ev_eq} with $\dot{\pfv} > 0$ and neglect all spatial derivatives of $d$ as we consider the spatially homogeneous solution. 
By performing integration over the thickness $h$, we get
\begin{eqnarray} 
\frac{\Gf h}{c_{\alpha}\lc} 
\der{\alpha(\pfv)}{\pfv}
= 2 (1-\pfv) 
\int_{-h/2}^{h/2}\psi^+_\text{e}\de z.
\label{S:ev_eq_h}
\end{eqnarray}

For simplicity, we limit our attention to PF-P model. To that end, we assume that the damage starts to evolve if the largest tensile stress reaches the tensile strength $\ft$. This allows us to write the integral on the right-hand side of~\Eref{S:ev_eq_h} in terms of $\ft$ as
\begin{align}
    \int_{-h/2}^{h/2}\psi^+_\text{e}\de z=\frac{\ft^2h}{12E}.
\label{eq:edb}
\end{align}
If we set $\pfv=0$ at the onset of cracking, Eqs.~\eqref{S:ev_eq_h} and \eqref{eq:edb} lead for $c_\alpha = 8/3$ and $\der{\alpha(\pfv)}{\pfv}=1$, recall~\Tref{tab:par_mod},
to the approximation of the length-scale parameter in the form
\begin{align}
    \lc  
    = 6 \left( \frac{3}{8}\frac{\Gf E}{\ft^2} \right).
    \label{E:lc}
\end{align}
The same result can be derived under some simplifying assumptions for the PF-B formulation,
\begin{align}
    \lc  
    = 6 \left( \frac{27}{256}\frac{\Gf E}{\ft^2} \right).
    \label{E:lc2}
\end{align}

\section{Numerical case study of a monolithic glass plate under bending} 
\label{S:NCS}

This first part of our study is devoted to monolithic glass plates only. This way, we want to separate the effect of polymer foils and focus purely on glass response. The response of laminated glass is discussed, and the numerical model is validated against experimental data in the next two sections.
The computational library FEniCS~\cite{logg2012automated} was used to implement the presented models for all numerical experiments; the source codes are available in~\cite{git_jarda}.
Other successful applications of the phase-field fracture in the framework of FEniCS can be found in~\cite{bleyer2017dynamic,natarajan2019phase}.

Our implementation strategy for the anisotropic and hybrid staggered approach is outlined in~Algorithms~\ref{alg:staggered} and \ref{alg:staggered_hyb}, respectively. For both algorithms, we update the phase-field parameter with the FEniCS-SNES solver, based on a semi-smooth Newton method for variational inequalities, to ensure that the parameter $\pfv$ cannot decrease in time and stays within the admissible $\left<0, 1\right>$ interval. Additionally, the iterations with a time-step are terminated using the energy convergence control with the tolerance of $\xi_\mathrm{SA}=10^{-6}$. Notice that the displacement sub-problem in the anisotropic staggered scheme (Algorithm~\ref{alg:staggered}) involves iterative energy minimisation using the Newton method with the relative tolerance $\xi_\mathrm{NM}=10^{-11}$, whereas the same step in Algorithm~\ref{alg:staggered_hyb} results in a linear problem, recall the discussion in~\Sref{S:hybrid}.

\begin{algorithm}[ht]
\SetAlgoLined
\KwData{}
set staggered approach tolerance $\xi_\mathrm{SA}$\;
set Newton method tolerance $\xi_\mathrm{NM}$\;
\For{$t\in\{t_0, \ldots, t_\mathrm{end}\}$} {
\While{$\xi^{(i)} > \xi_\mathrm{SA}$}{

get $\mathbf{u}_{t}^{(i)}$ by minimisation of $\mathcal{E}^{(i)}$ from \eqref{eq:EF} w.r.t. $\mathbf{u}$ using Newton method with tolerance $\xi_\mathrm{NM}$\;

get ${d}_{t}^{(i)}$ by iterative minimisation of $\mathcal{E}^{(i)}$ from \eqref{eq:EF} w.r.t. ${d}$\ using FEniCS-SNES solver\;
update $\mathcal{E}^{(i)}\leftarrow\mathcal{E}(\mathbf{u}_t^{(i)}, d_t^{(i)})$\;
update $\xi^{(i)}\leftarrow |\mathcal{E}^{(i)} - \mathcal{E}^{(i-1)}|/\mathcal{E}^{(i)}$\;
update $i\leftarrow i+1$\;
}
}
\caption{Anisotropic staggered approach}
\label{alg:staggered}
\end{algorithm}
\begin{algorithm}[ht]
\SetAlgoLined
\KwData{}
set staggered approach tolerance $\xi_\mathrm{SA}$\;
\For{$t\in\{t_0, \ldots, t_\mathrm{end}\}$} {
\While{$\xi^{(i)} > \xi_\mathrm{SA}$}{
get $\mathbf{u}_{t}^{(i)}$ from linear 
problem given by minimisation of $\mathcal{E}^{(i)}$ from \eqref{eq:EF_withoutsplit} w.r.t. $\mathbf{u}$ \;
get ${d}_{t}^{(i)}$ by minimisation of $\mathcal{E}^{(i)}$ from \eqref{eq:EF} w.r.t. ${d}$\ using FEniCS-SNES solver\;
update $\mathcal{E}^{(i)}\leftarrow\mathcal{E}(\mathbf{u}_t^{(i)}, d_t^{(i)})$\;
update $\xi^{(i)}\leftarrow |\mathcal{E}^{(i)} - \mathcal{E}^{(i-1)}|/\mathcal{E}^{(i)}$\;
update $i\leftarrow i+1$\;
}
}
\caption{Hybrid (isotropic-anisotropic) staggered approach}
\label{alg:staggered_hyb}
\end{algorithm}

\begin{figure}[ht]
     \centering
    \begin{tabular}{c c}
        \raisebox{-0.5\height}{\def\svgwidth{0.55\textwidth}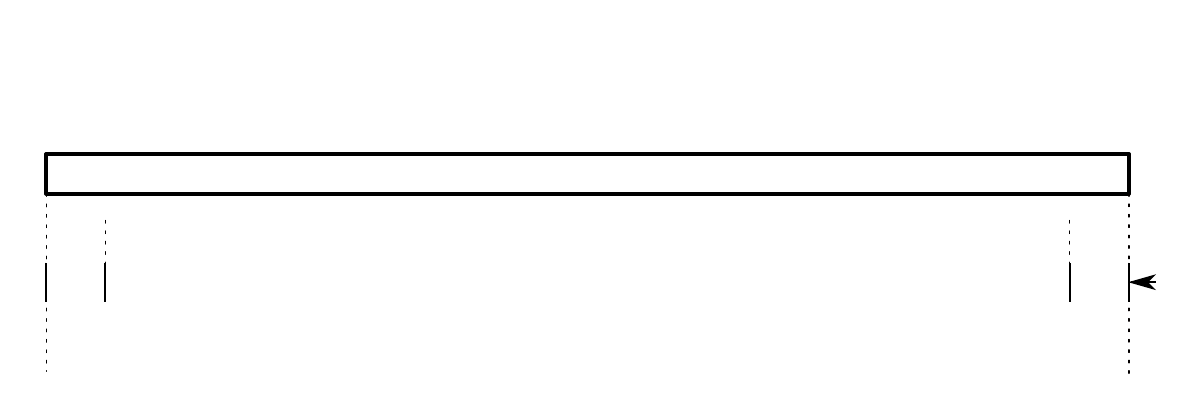} & 
        \raisebox{-0.5\height}{\def\svgwidth{0.39\textwidth}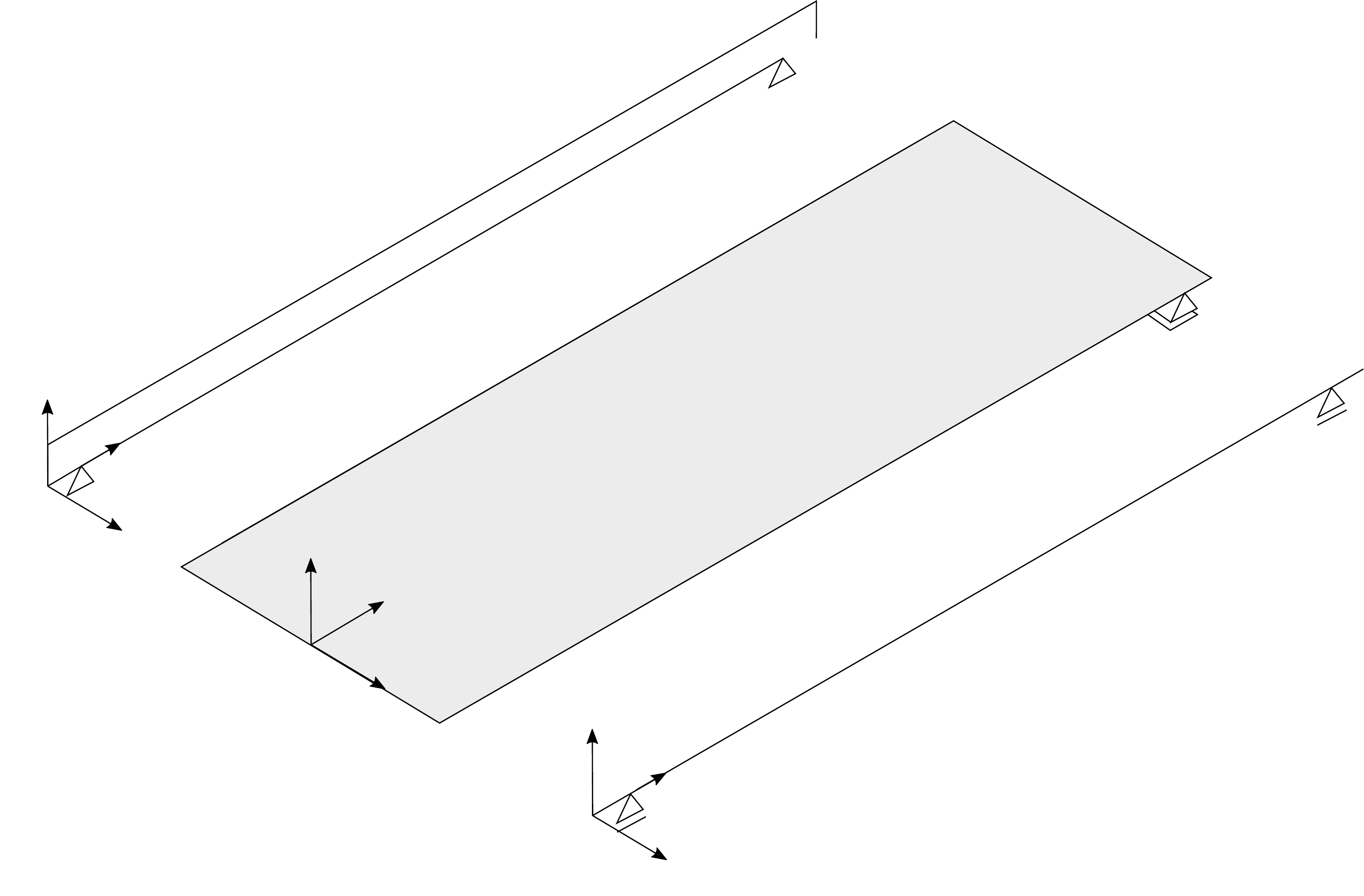} \\
    \end{tabular}
    \caption{Four-point bending loading scheme and geometry of a glass sample and three different spatially reduced models, i.e., 2D plane-stress (PS) longitudinal cross-section, Reissner-Mindlin plate (P), Timoshenko beam (B).}
    \label{fig:scheme_mon_models}
\end{figure}

\begin{figure}[ht]
    \centering
    \begin{tabular}{cc}
    \def\svgwidth{0.9\textwidth}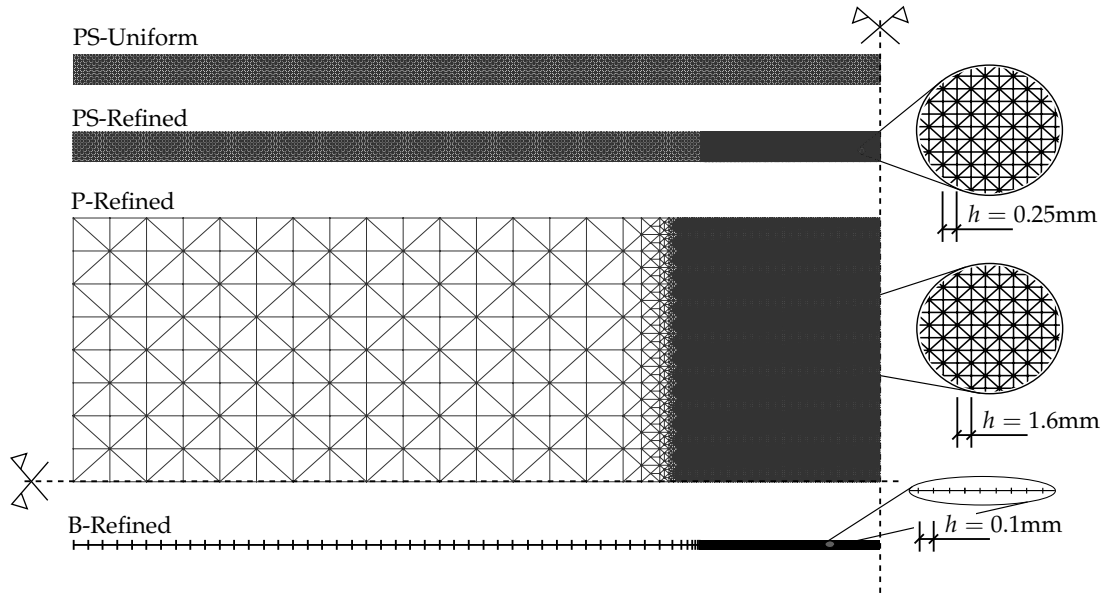
    \end{tabular}
    \caption{Regular uniform or refined discretization for the 2D plane-stress model (PS) and refined meshes used for a quarter of a plate (P) and a half of a beam (B).}
    \label{F:mesh}
\end{figure}

The numerical experiments were performed on a single glass sheet of dimensions 1,100~mm $\times$ 360~mm $\times$ 20~mm loaded in four-point bending, see~\Fref{fig:scheme_mon_models}. The prescribed displacement of loading head was gradually increased  until fracture with the loading-rate of 0.03~mm/s. The pseudo-time increment in Algorithms~\ref{alg:staggered} and \ref{alg:staggered_hyb} started initially at 0.1~s and was subsequently refined to 0.01~s and finally to 0.001~s close to localisation.
Regarding the discretization, we tested three different spatially reduced models schematically shown in~\Fref{fig:scheme_mon_models}, i.e.,
    2D plane-stress model (PS) of the longitudinal cross-section, 2D model using the Mindlin-Reissner plate theory (P), and
    1D model based on the Timoshenko beam theory (B).
For the simulations, a few types of meshes were tested, i.e.,
    PS-Uniform: regular uniform mesh with the element size of 2~mm, 
    PS-Refined mesh,
    P-Refined mesh or
    B-Refined mesh refined in the largest-bending-moment area,~\Fref{F:mesh}.
Let us highlight that all simulations reported in the following assume the symmetries shown in~\Fref{F:mesh}. Hence, the simulation results and the localised crack in particular must be interpreted by taking these symmetries into consideration.

Finite elements with linear basis functions were employed in this numerical study as the phase-field formulations require relatively fine meshes in any case. Prior to modelling damage, we tested the convergence of the solution to the linear elastic problem without glass fracture. For example for the plane-stress formulation, the errors in displacements and stresses were under 2.5\% for the uniform mesh and under 1.5\% for the refined variant, compared to the reference solution corresponding to the mesh of 1,100$\times$20 quadratic elements.

The length-scale parameter $\lc$ was set according to the element size, i.e., $\lc \approx 2 h_\text{min}$ with the smallest element size $h_\text{min}$, see~\cite{miehe2010thermodynamically}. Subsequently, the corresponding fracture energy $\Gf$ was derived according to~Eqs.~\eqref{eq:lcB} and ~\eqref{eq:lcP}, or according to the modified version for beams in~Eqs.~\eqref{E:lc} or~\eqref{E:lc2}.

\subsection{Effect of the phase-field formulation and mesh refinement}

\begin{figure}[!ht]
\centering
\begin{tabular}{c}
\includegraphics{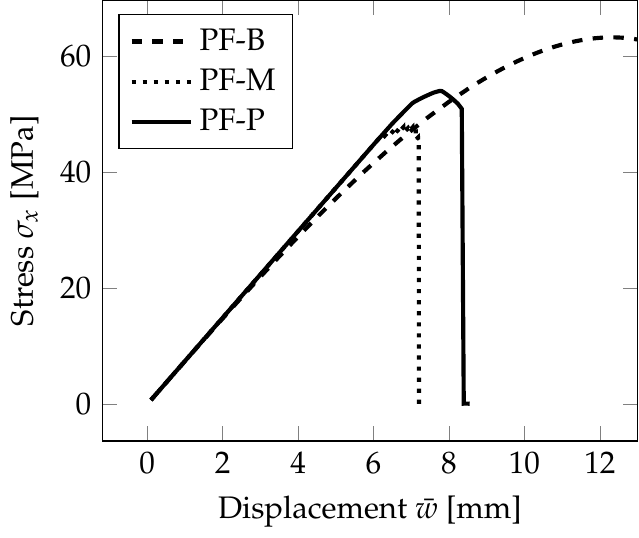}
\includegraphics{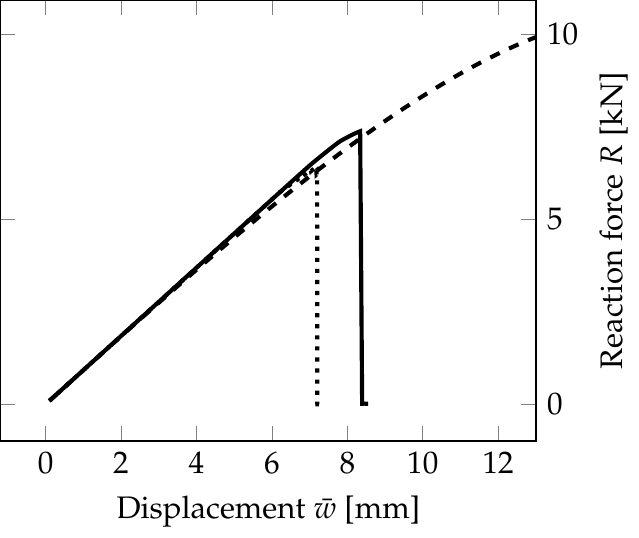}
\\
\def\svgwidth{100mm}
\begingroup%
  \makeatletter%
  \providecommand\color[2][]{%
    \errmessage{(Inkscape) Color is used for the text in Inkscape, but the package 'color.sty' is not loaded}%
    \renewcommand\color[2][]{}%
  }%
  \providecommand\transparent[1]{%
    \errmessage{(Inkscape) Transparency is used (non-zero) for the text in Inkscape, but the package 'transparent.sty' is not loaded}%
    \renewcommand\transparent[1]{}%
  }%
  \providecommand\rotatebox[2]{#2}%
  \newcommand*\fsize{\dimexpr\f@size pt\relax}%
  \newcommand*\lineheight[1]{\fontsize{\fsize}{#1\fsize}\selectfont}%
  \ifx\svgwidth\undefined%
    \setlength{\unitlength}{744.35676074bp}%
    \ifx\svgscale\undefined%
      \relax%
    \else%
      \setlength{\unitlength}{\unitlength * \real{\svgscale}}%
    \fi%
  \else%
    \setlength{\unitlength}{\svgwidth}%
  \fi%
  \global\let\svgwidth\undefined%
  \global\let\svgscale\undefined%
  \makeatother%
  \begin{picture}(1,0.35896187)%
    \lineheight{1}%
    \setlength\tabcolsep{0pt}%
    \put(0,0){\includegraphics[width=\unitlength,page=1]{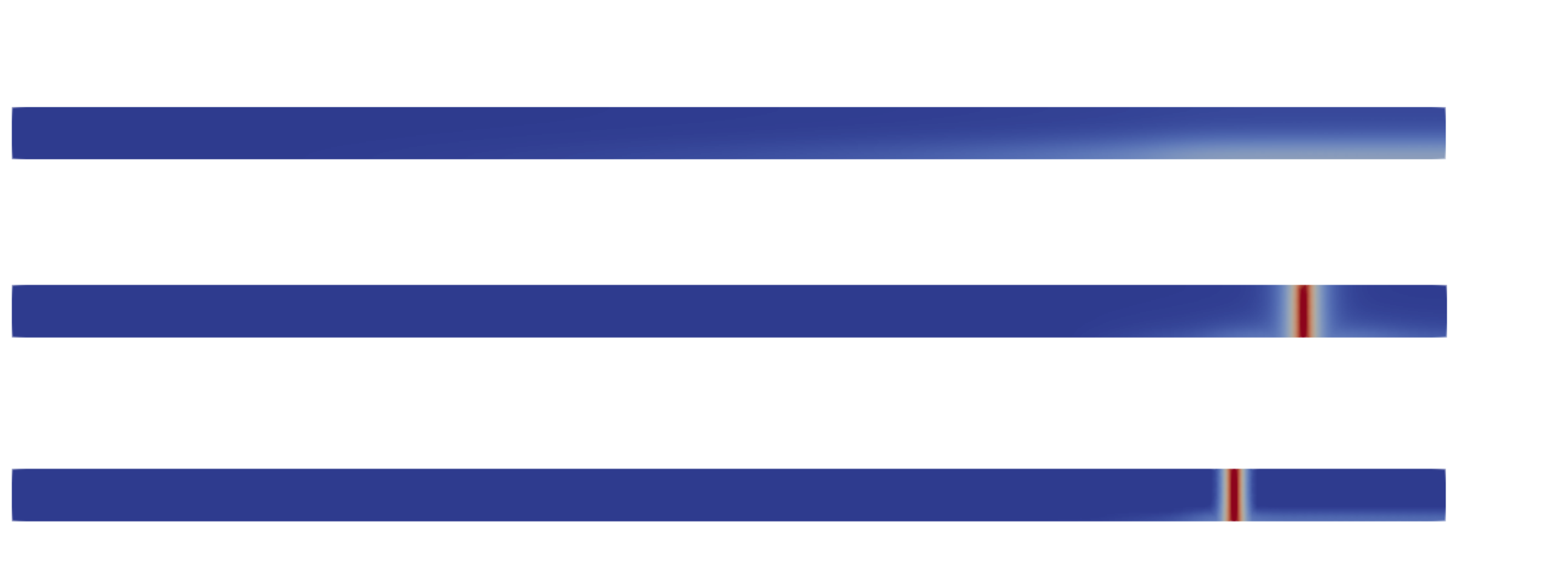}}%
    \put(0.00504493,0.30348848){\color[rgb]{0,0,0}\makebox(0,0)[lt]{\lineheight{1.25}\smash{\begin{tabular}[t]{l}PF-B, $\bar{w}=$ 15.0 mm\end{tabular}}}}%
    \put(0.00185537,0.18634668){\color[rgb]{0,0,0}\makebox(0,0)[lt]{\lineheight{1.25}\smash{\begin{tabular}[t]{l}PF-M, $\bar{w}=$ 7.2 mm\end{tabular}}}}%
    \put(0.00198413,0.07422167){\color[rgb]{0,0,0}\makebox(0,0)[lt]{\lineheight{1.25}\smash{\begin{tabular}[t]{l}PF-P, $\bar{w}=$ 8.39 mm\end{tabular}}}}%
    \put(0,0){\includegraphics[width=\unitlength,page=2]{figure05c.pdf}}%
  \end{picture}%
\endgroup%

\end{tabular}
\caption{PS-Uniform mesh, anisotropic staggered solver with the spectral-decomposition split: Comparison of phase-field formulations in terms of the evolution of the largest tensile stress at the midpoint and the overall reaction under the loading points for the prescribed displacement, complemented with the damage evolution plot showing the position of the localised cracks.
}
\label{F:Comp_F_U}
\end{figure}

\begin{figure}[!ht]
\centering
\begin{tabular}{c}
\vspace{0mm}\\
\includegraphics{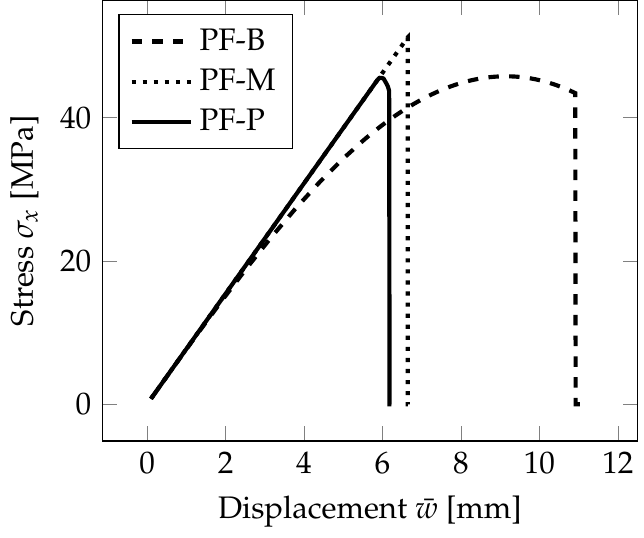}
\includegraphics{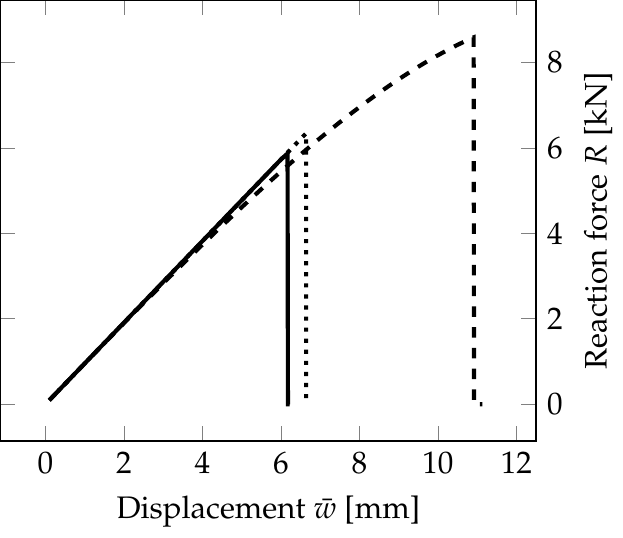}
\\
\def\svgwidth{100mm}
\begingroup%
  \makeatletter%
  \providecommand\color[2][]{%
    \errmessage{(Inkscape) Color is used for the text in Inkscape, but the package 'color.sty' is not loaded}%
    \renewcommand\color[2][]{}%
  }%
  \providecommand\transparent[1]{%
    \errmessage{(Inkscape) Transparency is used (non-zero) for the text in Inkscape, but the package 'transparent.sty' is not loaded}%
    \renewcommand\transparent[1]{}%
  }%
  \providecommand\rotatebox[2]{#2}%
  \newcommand*\fsize{\dimexpr\f@size pt\relax}%
  \newcommand*\lineheight[1]{\fontsize{\fsize}{#1\fsize}\selectfont}%
  \ifx\svgwidth\undefined%
    \setlength{\unitlength}{726.97515693bp}%
    \ifx\svgscale\undefined%
      \relax%
    \else%
      \setlength{\unitlength}{\unitlength * \real{\svgscale}}%
    \fi%
  \else%
    \setlength{\unitlength}{\svgwidth}%
  \fi%
  \global\let\svgwidth\undefined%
  \global\let\svgscale\undefined%
  \makeatother%
  \begin{picture}(1,0.36828201)%
    \lineheight{1}%
    \setlength\tabcolsep{0pt}%
    \put(0,0){\includegraphics[width=\unitlength,page=1]{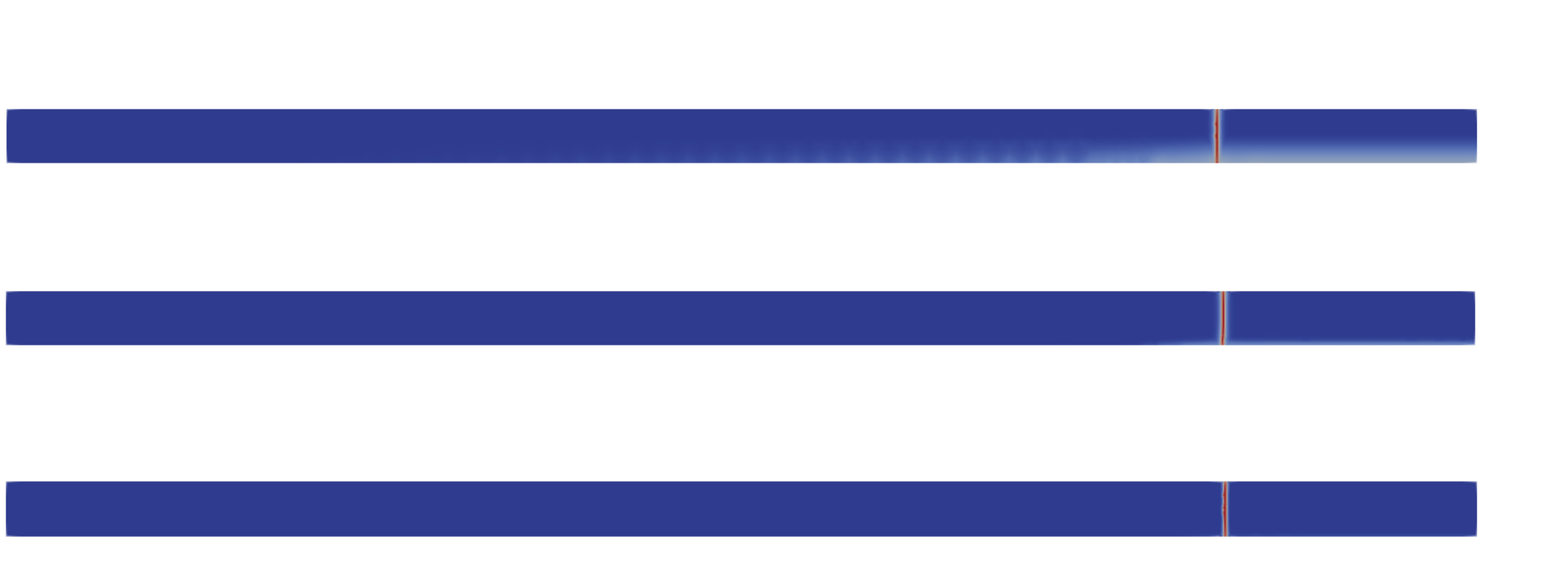}}%
    \put(0.00224413,0.31148226){\color[rgb]{0,0,0}\makebox(0,0)[lt]{\lineheight{1.25}\smash{\begin{tabular}[t]{l}PF-B, $\bar{w}=11.6$ mm\end{tabular}}}}%
    \put(-0.00102169,0.19153968){\color[rgb]{0,0,0}\makebox(0,0)[lt]{\lineheight{1.25}\smash{\begin{tabular}[t]{l}PF-M, $\bar{w}=7.05$ mm\end{tabular}}}}%
    \put(-0.00088985,0.07673382){\color[rgb]{0,0,0}\makebox(0,0)[lt]{\lineheight{1.25}\smash{\begin{tabular}[t]{l}PF-P, $\bar{w}=6.26$ mm\end{tabular}}}}%
    \put(0,0){\includegraphics[width=\unitlength,page=2]{figure06c.pdf}}%
  \end{picture}%
\endgroup%

\end{tabular}
\caption{PS-Refined mesh, anisotropic staggered solver with the spectral-decomposition split: Comparison of phase-field formulations in terms of the evolution of the largest tensile stress at the midpoint and the overall reaction under the loading points for the prescribed displacement, complemented with the damage evolution plot showing the position of the localised cracks.}
\label{F:Comp_F_R}
\end{figure}

First, we present a comparison of three responses corresponding to the individual phase-field formulations using the PS model. In~Figures~\ref{F:Comp_F_U} and \ref{F:Comp_F_R}, the diagrams show the evolution of the largest tensile stress $\sigma_x$ at the midpoint and the overall reaction $R$ under the loading points for the prescribed vertical displacement of the loading head $\bar{w}$. 
It can be seen in~\Fref{F:Comp_F_U} that the tensile stress evolutions differ quite significantly for the regular mesh. The PF-B model yields a nonlinear stress-strain response
from the beginning of the loading test, so no initial linear phase is visible in the plots in~\Fref{F:Comp_F_U}. Moreover, the failure stress differs for the three formulations.
We attribute this discrepancy to a relatively coarse mesh 
and
linear basis functions 
for the displacement field yielding constant stresses and strain energy density within an element. 
Consequently, 
the fracture occurs at different prescribed loading levels.

{}
Therefore, we refined the mesh in the area of the largest bending moment, where the cracks are supposed to initiate. The element size in this area is about 0.25~mm, see~\Fref{F:mesh}. The evolution of the normal stress and reaction and the damage after the crack localisation are shown in~\Fref{F:Comp_F_R}. In this case, the failure stresses are closer to each other and to the adopted tensile strength of glass. We assume that for even finer mesh, the performance of the PF-M and PF-P models would be almost indistinguishable. On the other hand, a significant 
nonlinear response prior to
fracture can be seen for the PF-B formulation.
Also the crack localised in this particular case for an almost two-times higher prescribed displacement.

\begin{figure}[!ht]
    \centering
    \def\svgwidth{110mm}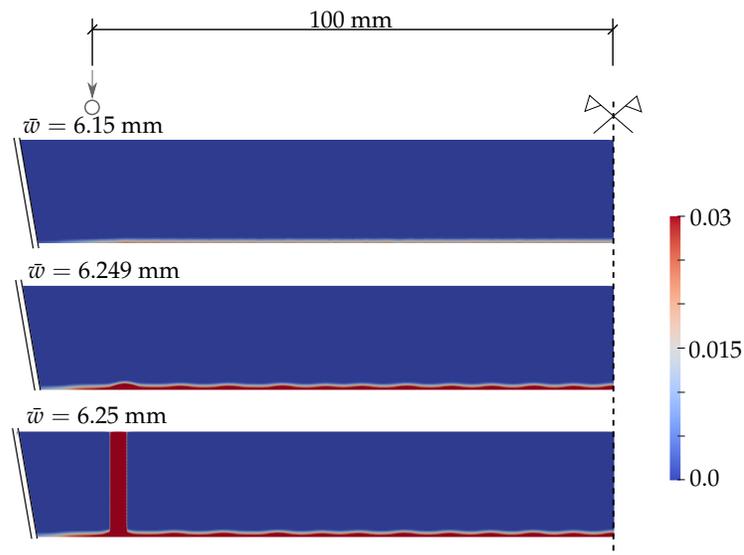
    \caption{PS-Refined mesh, anisotropic staggered solver with the spectral-decomposition split: Evolution of the phase-field parameter using PF-P formulation.}
    \label{fig:dam_ev}
\end{figure}

Further, \Fref{fig:dam_ev} shows the phase-field parameter evolution before the damage is localised in one crack for the PF-P formulation. We use one scale for all the plots. Initially, the damage starts to evolve in the area of the largest bending moment. Then, a few relatively equidistant 
short ridges 
appear, and finally, one crack, corresponding to the maximum of the phase-field variable,
localises close to the loading point. 
Hence, the phase-field model seems to predict the initiation of multiple regularly spaced cracks, from whose only one localises, as the localised solution corresponds to a lower energy value, e.g.,~\cite{hunt1995principles,brocca1997analysis,audy1993localization}. 

Based on this comparison, we decided to prefer the PF-P formulation for the next analyses as it provides the initial linear elastic response of glass and preserves the variational structure of the problem. The refined mesh discretisations were used for all following examples in this numerical study and all three spatially reduced models; the label ``Refined'' is omitted to shorten the notation. 

\subsection{Effect of the type of the solver and tension/compression energy split}

Subsequently, we briefly compare the anisotropic and hybrid formulation,~\Sref{S:hybrid}, within the staggered scheme for our example, combined with two different ways for the tension-compression split of the strain tensor from \Sref{S:anis},
i.e., the volumetric-deviatoric split and the spectral decomposition. 
\begin{figure}[ht]
\centering
\includegraphics{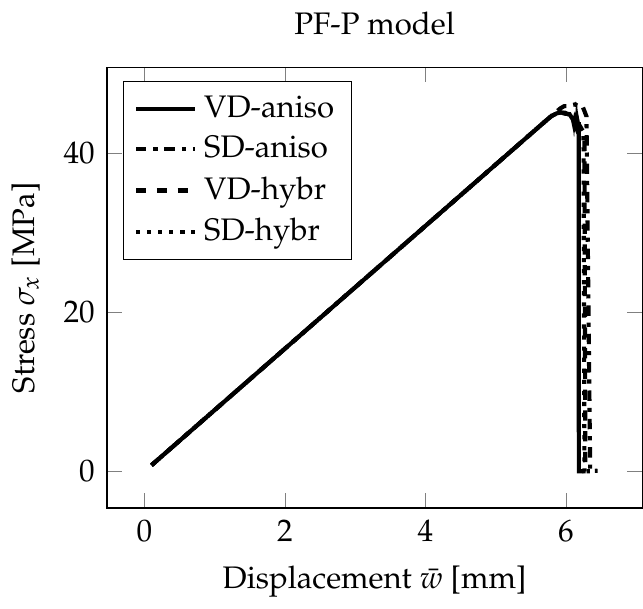}
\includegraphics{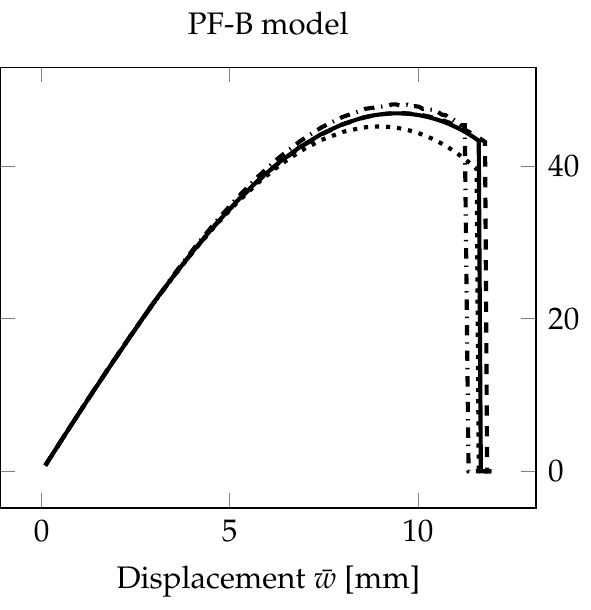}
\caption{PS-Refined mesh: Comparison of the anisotropic solver, Algorithm~\ref{alg:staggered}, employing the volumetric-deviatoric split (VD-aniso) or spectral decomposition approach (SD-aniso) with their counterparts using a linear hybrid formulation (VD-hybr and SD-hybr), Algorithm~\ref{alg:staggered_hyb}.}
\label{F:split}
\end{figure}
As can be seen from \Fref{F:split}, both ways of the formulation and decomposition deliver almost identical response for the PF-P model of a thin glass plate under bending. For completeness, we show the same comparison also for the PF-B formulation. The observed differences can be attributed to the predicted early-stage degradation.
However, the effect of the choice of the fully anisotropic or hybrid formulation and the split is negligible for our purposes.

\subsection{Effect of the dimensional reduction}
\label{S:comp_dim}

Before comparing the fracture response corresponding to different dimensional reductions, we illustrate the effect of the relationship between the fracture energy $\Gf$ set according to Eqs.~\eqref{eq:lcP} and \eqref{E:lc} for the selected $\lc$ and $\ft$.
\begin{figure}[ht]
\centering
\includegraphics{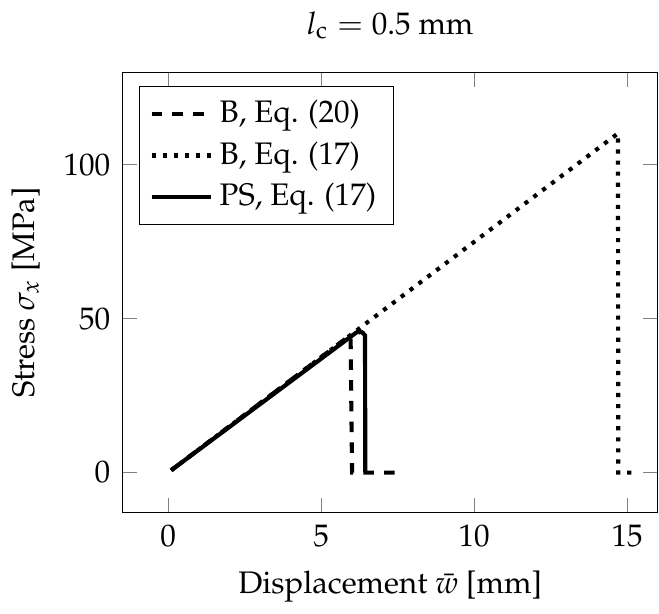}
\includegraphics{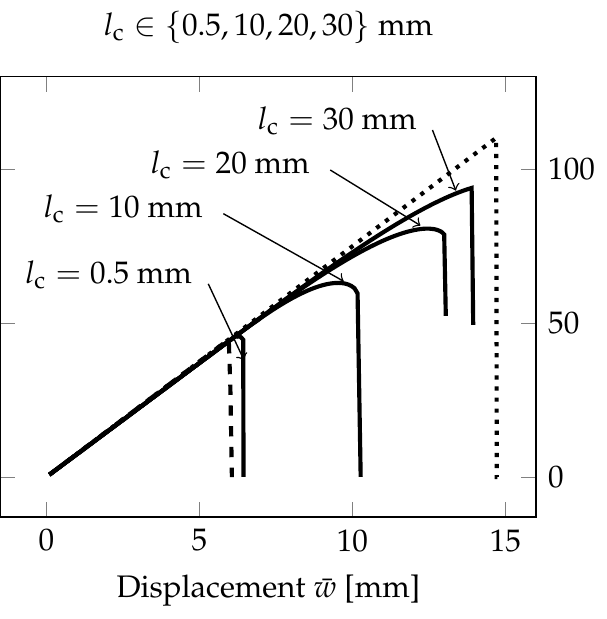}
    \caption{PF-P model, anisotropic staggered approach with the spectral-decomposition split: Comparison of the plane-stress model (PS) with the formulation derived for beams (B) using two constitutive relationships for $G_\mathrm{c}$, $l_\mathrm{c}$, and $f_\mathrm{t}$, i.e., ${\Gf}/{\lc} = {8\ft^2}/(3E)$ from~\Eref{eq:lcP} and ${6\Gf}/{\lc} = {8\ft^2}/{(3E)}$ from~\Eref{E:lc}.}
\label{fig:beam_comparison}
\end{figure}

\Fref{fig:beam_comparison} shows the evolution of the largest tensile stress at the mid-point of the span for the increasing prescribed deflection. The PF-P formulation with the spectral decomposition was used within the staggered solver. The derived fracture energy $\Gf$ corresponds to the critical stress $\ft=45$~MPa, and the length-scale parameter for the PS model set to $\lc=0.025h = 0.5$~mm, i.e., twice the element size, or $\lc\in\{10,20,30\}$~mm, i.e., $\lc\in\{0.5h,h,1.5h\}$. For the B model, the length-scale parameter was fixed to the value $\lc= 0.2$~mm, i.e., twice the element size, as its change does not affect the results in contrast with the PS model.

For the beam formulation with the mesh density shown in~\Fref{F:mesh}, the largest failure stress at the bottom surface under tension is more than doubled compared to the given critical stress $\ft$ with $\lc$ provided by~\Eref{eq:lcP}.
The adjusted fracture energy according to~\Eref{E:lc} provides the response comparable to the plane-stress formulation, and the fracture occurs close to the prescribed tensile strength if the length-scale parameter is much smaller than the thickness of the glass layer, e.g., $\lc=0.025h = 0.5$~mm, see \Fref{fig:beam_comparison}. 
On the other hand, if the crack is more diffused, the failure stress on the bottom surface increases. Then,
a nonlinear response
can also be seen in the stress-displacement diagram for the PS model. If the length-scale parameter is greater than the glass thickness,
the response of the PS model is approaching the upper bound given by the beam theory using the fracture energy derived from~\Eref{eq:lcP}.

\begin{figure}[!ht]
\centering
\includegraphics{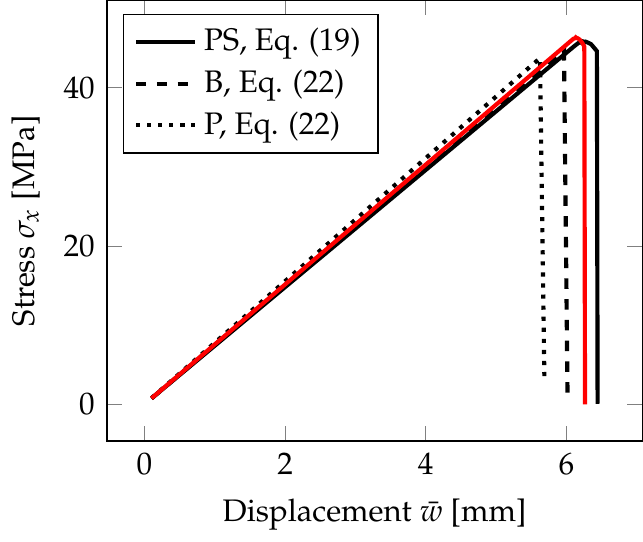}
\includegraphics{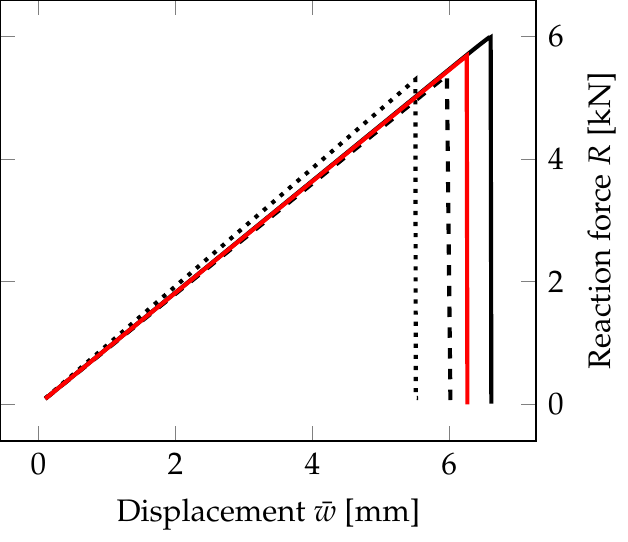}
\def\svgwidth{110mm}
\begingroup%
  \makeatletter%
  \providecommand\color[2][]{%
    \errmessage{(Inkscape) Color is used for the text in Inkscape, but the package 'color.sty' is not loaded}%
    \renewcommand\color[2][]{}%
  }%
  \providecommand\transparent[1]{%
    \errmessage{(Inkscape) Transparency is used (non-zero) for the text in Inkscape, but the package 'transparent.sty' is not loaded}%
    \renewcommand\transparent[1]{}%
  }%
  \providecommand\rotatebox[2]{#2}%
  \newcommand*\fsize{\dimexpr\f@size pt\relax}%
  \newcommand*\lineheight[1]{\fontsize{\fsize}{#1\fsize}\selectfont}%
  \ifx\svgwidth\undefined%
    \setlength{\unitlength}{801.32567858bp}%
    \ifx\svgscale\undefined%
      \relax%
    \else%
      \setlength{\unitlength}{\unitlength * \real{\svgscale}}%
    \fi%
  \else%
    \setlength{\unitlength}{\svgwidth}%
  \fi%
  \global\let\svgwidth\undefined%
  \global\let\svgscale\undefined%
  \makeatother%
  \begin{picture}(1,0.57692372)%
    \lineheight{1}%
    \setlength\tabcolsep{0pt}%
    \put(0,0){\includegraphics[width=\unitlength,page=1]{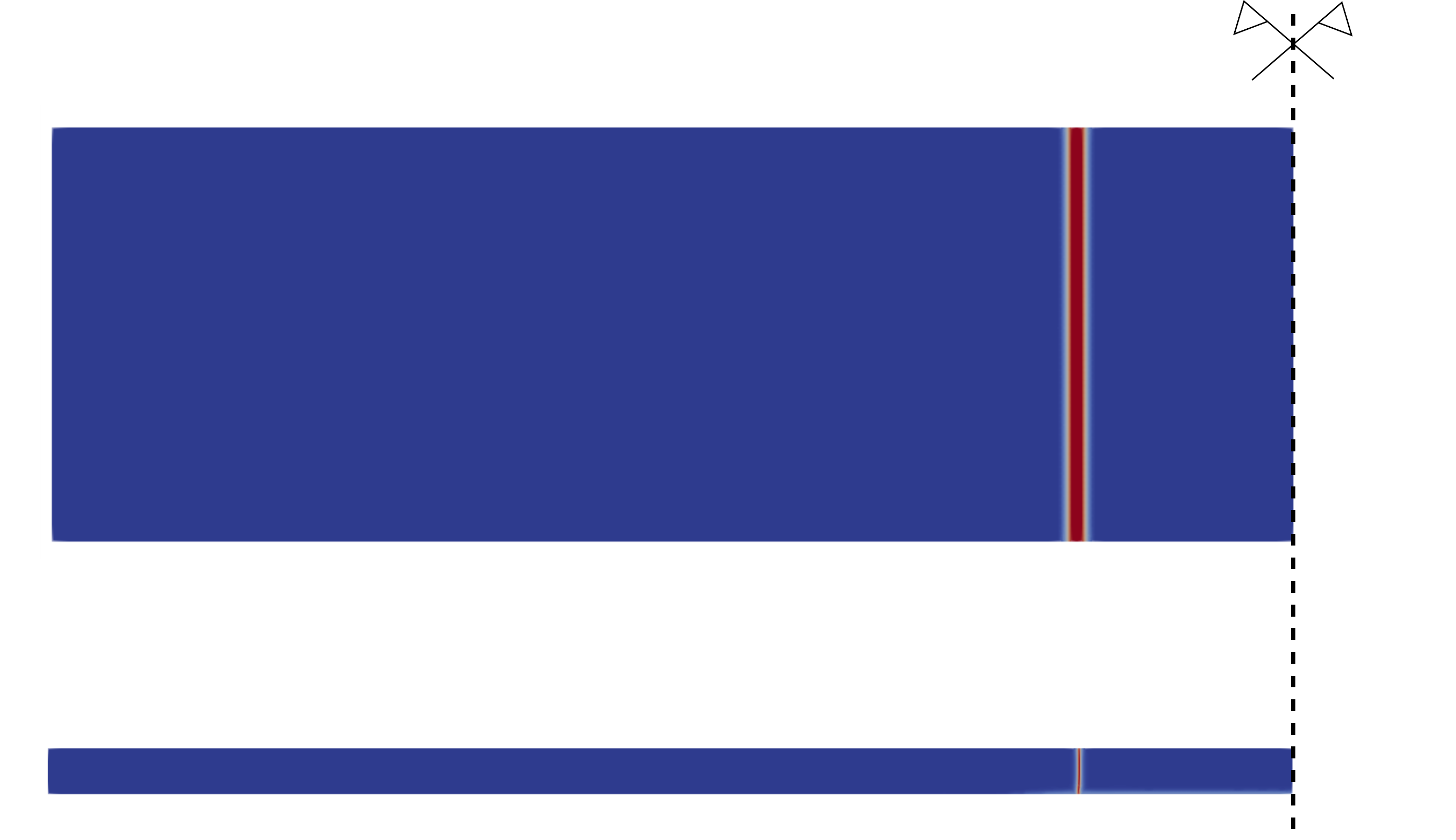}}%
    \put(0.03114438,0.13381815){\color[rgb]{0,0,0}\makebox(0,0)[lt]{\lineheight{1.25}\smash{\begin{tabular}[t]{l}Beam\end{tabular}}}}%
    \put(0,0){\includegraphics[width=\unitlength,page=2]{figure10c.pdf}}%
    \put(0.03192794,0.50601303){\color[rgb]{0,0,0}\makebox(0,0)[lt]{\lineheight{1.25}\smash{\begin{tabular}[t]{l}Plate\end{tabular}}}}%
    \put(0,0){\includegraphics[width=\unitlength,page=3]{figure10c.pdf}}%
    \put(0.03016334,0.0663136){\color[rgb]{0,0,0}\makebox(0,0)[lt]{\lineheight{1.25}\smash{\begin{tabular}[t]{l}Plane stress\end{tabular}}}}%
    \put(0,0){\includegraphics[width=\unitlength,page=4]{figure10c.pdf}}%
  \end{picture}%
\endgroup%

\caption{PF-P formulation, anisotropic staggered approach with the spectral-decomposition split: Comparison of plane stress (PS) {for the PS-Refined mesh from~\Fref{F:mesh} (black line) or additionally refined in the area of the expected crack propagation (red line)}, beam (B), and plate (P) model in terms of the evolution of the largest tensile stress at the midpoint, the overall reaction, and the damage evolution showing the position of the localised cracks.}
\label{F:lc}
\end{figure}

This comparison illustrates that the fracture energy has to be set with respect to the applied dimensional reduction, loading type, and the value of the length-scale parameter compared with the thickness of the structure under bending. For example in~\cite{kiendl2016phase}, the authors report the same response of their phase-field fracture model for plates and the reference solution for a solid using a constant $\Gf$ value and the length-scale parameter $\lc$ equal to the plate thickness. However, this identical cannot be achieved for any values of the length-scale parameter.

\Fref{F:lc} compares the response of the PS, B, and P model for a glass monolith simply supported on two sides.
The evolution of the largest tensile stress and the overall reaction force $R$ under the loading points for the prescribed deflection is plotted for the three models.
The failure stress is almost the same for the beam and plate formulations and slightly higher for the PS model. The stiffnesses of the glass monolith corresponding to the beam theory and the PS model are equivalent, but the fracture occurs later for the PS model. The red line corresponds to a mesh additionally refined in the area of the expected crack propagation to illustrate how the responses for the B a PS models converge if the PS mesh is refined.

On the other hand, the plate formulation is stiffer and the glass fractured for a lower prescribed deflection. We attribute these small differences again to a rather coarse mesh for the plate in some unfractured regions and to linear basis functions used in this analysis. We expect the differences to decrease at the cost of higher computational demands.
The crack appeared at the same position near the loading  for the PS and P model, see~\Fref{F:lc}. For the B model, the position of fracture is different, but still at the region of the constant largest bending moment.

\begin{figure}[ht]
\centering
\includegraphics{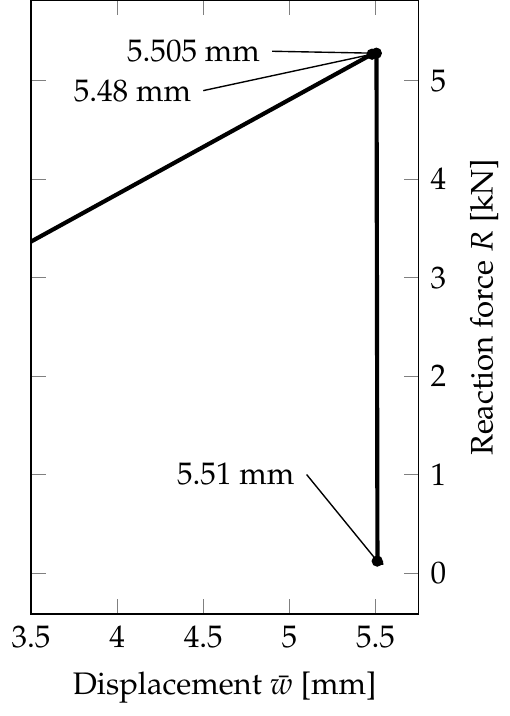}
\def\svgwidth{90mm}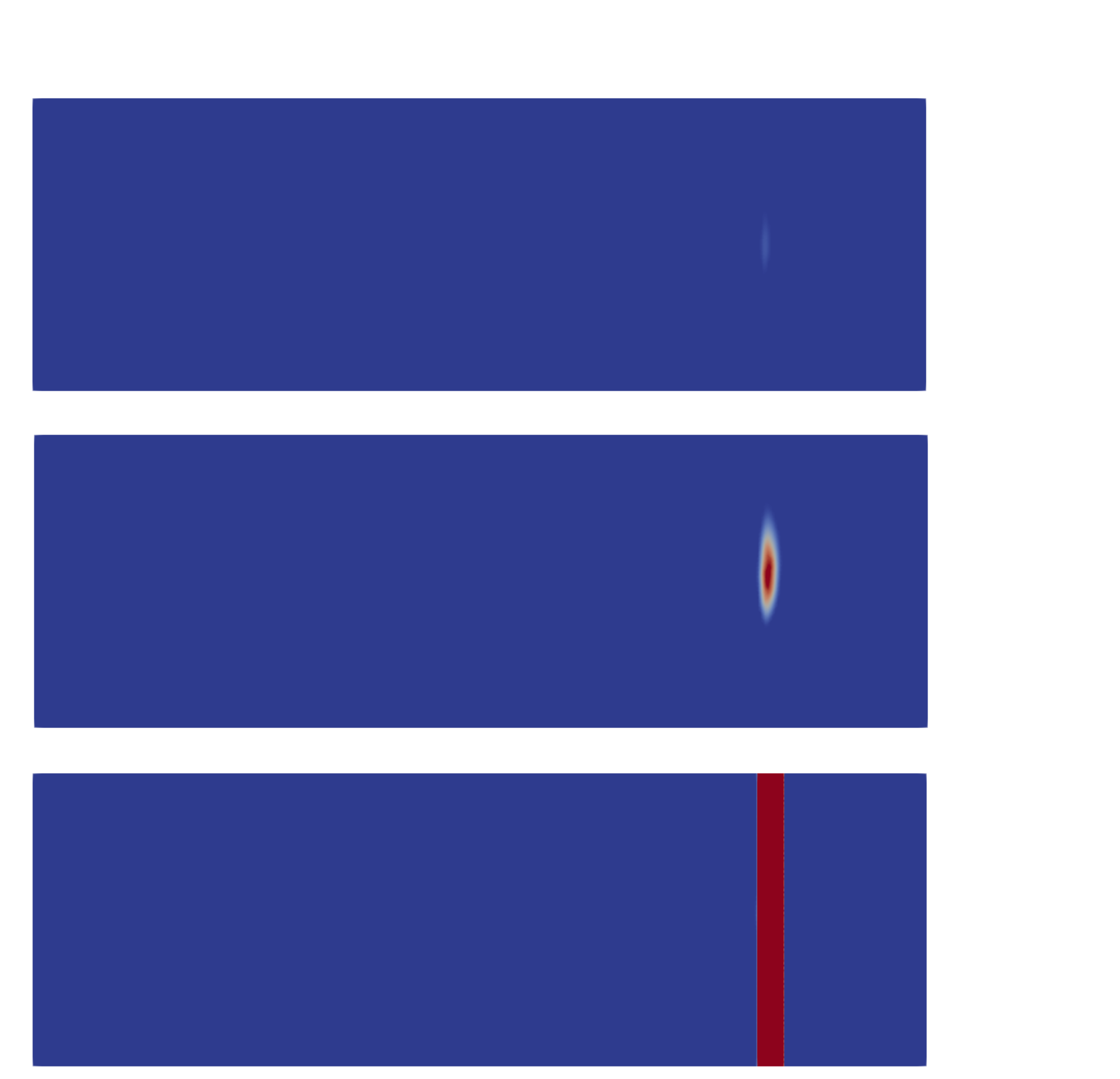
\caption{PF-P formulation, anisotropic staggered approach with the spectral-decomposition split: Evolution of the phase-field parameter on a quarter of the glass plate.}
\label{F:pl_fr_quar}
\end{figure}

For the plate model, the evolution of the phase-field parameter is displayed in~\Fref{F:pl_fr_quar}. Each rectangle represents a quarter of the glass plate for a different magnitude of the prescribed loading.
The damage starts to initiate at about a quarter of the plate width near the loading point and subsequently localises into a straight crack.

\subsection{Reduced glass strength near plate edges}

\begin{figure}[ht]
    \centering
    \def\svgwidth{\textwidth}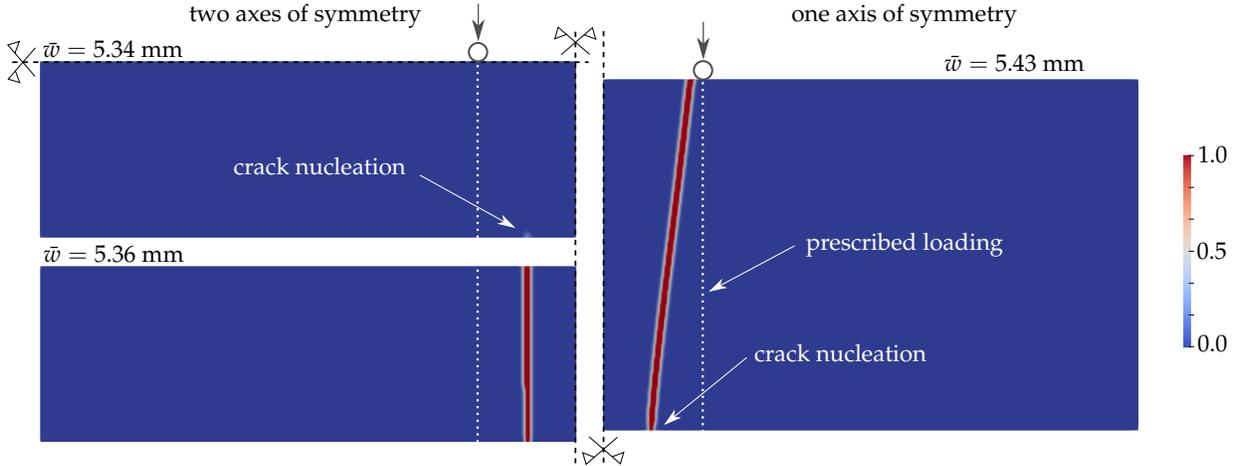
    \caption{PF-P formulation, P mesh, anisotropic staggered approach with the spectral-decomposition split: Evolution of the phase-field parameter on a quarter of the glass plate and the final crack pattern on a half of the plate.}
    \label{fig:PFev_plate}
\end{figure}

The tensile strength of thin glass plates is highly affected by micro-defects and scratches induced during the production, transport, and handling. Moreover, the edge strength depends on the quality of edge finishing. Because the region of the largest tensile stresses includes parts of both edges for the four-point bending tests, the crack mostly initiates from a spot on edge. 
As suggested for example in the German glass standard~\cite{din2010glas}, the strength on edges can be reduced to initiate the cracking from the edge.
Near the bottom edge about 50~mm from the midspan, we modified the strength to 80\% of the given tensile strength in a $1.5\lc\times1.5\lc$ square, i.e., $4.8\text{~mm}\times4.8\text{~mm}$ in our example.  
Then, the crack starts to initiate from this predefined area toward the opposite edge,
as can be seen in~\Fref{fig:PFev_plate}. For a smaller area, e.g. $\lc\times\lc$, the localisation does not occur from an edge point. 
If we use a quarter of the plate with two axes of symmetry, the final crack is almost perpendicular to the long edges as the strength was reduced on two opposite sides. On the other hand, the crack is inclined when one half of the laminated glass plate is used in simulation,~\Fref{fig:PFev_plate}.  

Similarly, the strength can be reduced along the whole edge. The lower strength has to be assigned not only to the nodes directly on the edge but for a band of nodes to overcome the cracking initiated on the inner surface. However, we are unable to reproduce the typical V-shape fracture patterns, shown in~\Fref{fig:fract_1st} on page~\pageref{fig:fract_1st}, with the quasi-static plate model considering homogeneous material data in the plate interior; only one crack resulted from all numerical simulations. 

\section{Experimental testing on laminated glass}
\label{S:ETS}

The material parameters needed for the numerical model were obtained from both the literature review and experimental testing on three-layer laminated glass samples presented in this section.

\subsection{Material composition of laminated glass}

Two types of laminated glass samples were tested under quasi-static loading. All of them were three-layer plates with a polymer interlayer and two glass layers of the same type. The material specification is summarised in~\Tref{tab:samples}. The nominal dimensions of samples were as follows: the length $l = 1.1$~m, the width $b = 0.36$~m, and the thicknesses of layers $h_1/h_2/h_3 = 10/0.76/10$~mm.
\begin{table}[ht]
\centering{
\begin{tabular}{cccc}
\hline
name & glass & interlayer & samples\\
\hline 
ANG-EVA & annealed & ethylen-vinyl acetate & 5 
\\
ANG-PVB & annealed & polyvinyl butyral & 5
\\
\hline
\end{tabular}
}
\caption{Material composition of layers in tested specimens.}
\label{tab:samples}
\end{table}

\subsection{Testing of polymers and the material model}
\label{S:MPF}

In the present study, we exploit the results of our extensive experimental program reported in~\cite{hana2019experimental}. 
Therein, small three-layer cylindrical samples were loaded in dynamic torsion or dynamic single-lap shear under different ambient temperatures and applied frequencies. Subsequently, a set of parameters of the generalised Maxwell model for each interlayer were identified.
For both polymer foils, the series employed in 
the numerical modelling are summarised in~\Tref{tab:maxwell}.

\begin{table}[ht]
\centering
\begin{tabular}{lrrrlrr}
\hline
& {EVA} & {PVB} & \hspace{2cm} & & {EVA} & {PVB}\\
$\tau_p$ [s] & $G_p$ [kPa] & $G_p$ [kPa] && $\tau_p$ [s] & $G_p$ [kPa] & $G_p$ [kPa] \\
\hline
$10^{-9}$ & 6,933.9 & -- && $10^{2}$ & 445.1  & 587.2 \\
$10^{-8}$ & 3,898.6 & -- && $10^{3}$ & 300.1  & 258.0 \\
$10^{-7}$ & 2,289.2 & -- && $10^{4}$ & 401.6 & 63.8 \\
$10^{-6}$ & 1,672.7 & -- && $10^{5}$ & 348.1 & 168.4\\
$10^{-5}$ & 761.6  & 1,782,124.2 && $10^{6}$ & 111.6 & -- \\
$10^{-4}$ & 2,401.0 & 519,208.7 && $10^{7}$ & 127.2 & -- \\
$10^{-3}$ & 65.2   & 546,176.8 && $10^{8}$ & 137.8 & -- \\
$10^{-2}$ & 248.0  & 216,893.2 && $10^{9}$ & 50.5 & -- \\
$10^{-1}$ & 575.6  & 13,618.3 && $10^{10}$ & 322.9 & -- \\
$10^{0}$ & 56.3    & 4,988.3 && $10^{11}$ & 100.0 & -- \\
$10^{1}$ & 188.6   & 1,663.8 && $10^{12}$ & 199.9 & --\\
\hline
\end{tabular}
\caption{Prony series for the generalised Maxwell model with the relaxation times $\tau_p$ and corresponding shear moduli $G_p$ for the reference temperature $T_0=20\,^\circ\mathrm{C}$ and the long-term moduli $G_\infty=682.18$~kPa for EVALAM 80-120 (EVA) and $G_\infty=232.26$~kPa for TROSIFOL~ BG~R20 (PVB), \cite{hana2019experimental}.}
\label{tab:maxwell}
\end{table}

To simplify the formulation, we assume that the 
time/temperature-dependent
response of the interlayer can be approximated by an equivalent elastic material with the shear modulus $G$ in the middle of each time interval $t$ for the temperature $T$, see e.g.~\cite{Duser:1999:AGBL}. Then, the shear modulus of the interlayer is evaluated in each time instant according to 
\begin{equation}
    G(t,T)
    \approx
    G_\infty+\sum_{p=1}^{P}G_p \exp^\frac{- t/2}{a_T(T)\tau_p},
    \label{E:G_tT}
\end{equation}
with the Prony series $(G_p, \tau_p)_{p=1..P}$ and the long-term shear modulus $G_\infty$ from \Tref{tab:maxwell}.
For the shift parameter $a_{T}$ reflecting the temperature-dependency of the polymer interlayers,
we employed the  Williams-Landel-Ferry equation \cite{christensen2012theory}
\begin{equation}
\log a_{T}(T)=\frac{-C_1(T-T_0)}{C_2+T-T_0}.
\end{equation}
The model parameters $C_1$ and $C_2$ associated with the reference temperature $T_0$ are listed in~\Tref{tab:WLF}. 
\begin{table}[ht]
    \centering
\begin{tabular}{lcccc}
	\hline
	& & EVA 	& PVB	&	\\
	\hline
Reference temperature & $T_0$ & 20 & 20 & $^\circ\mathrm{C}$  \\
Parameters	&  $C_1$ & 339.102	&	8.635 & -- \\
& $C_2$ &	1,185.816 &	42.422 &  $^\circ\mathrm{C}$\\
Shift parameter & $\alpha_T(T=25^\circ\mathrm{C})$ & 0.03769 & 0.1229  & -- \\
	\hline
\end{tabular}
    \caption{Parameters for the time-temperature superposition using the William-Landel-Ferry equation~\cite{christensen2012theory}.}
    \label{tab:WLF}
\end{table}

\subsection{Quasi-static bending tests}
\label{S:QSB}

\begin{figure}[!ht]
\centerline{
\begin{tabular}{cc}
\multicolumn{2}{c}{\def\svgwidth{130mm}
	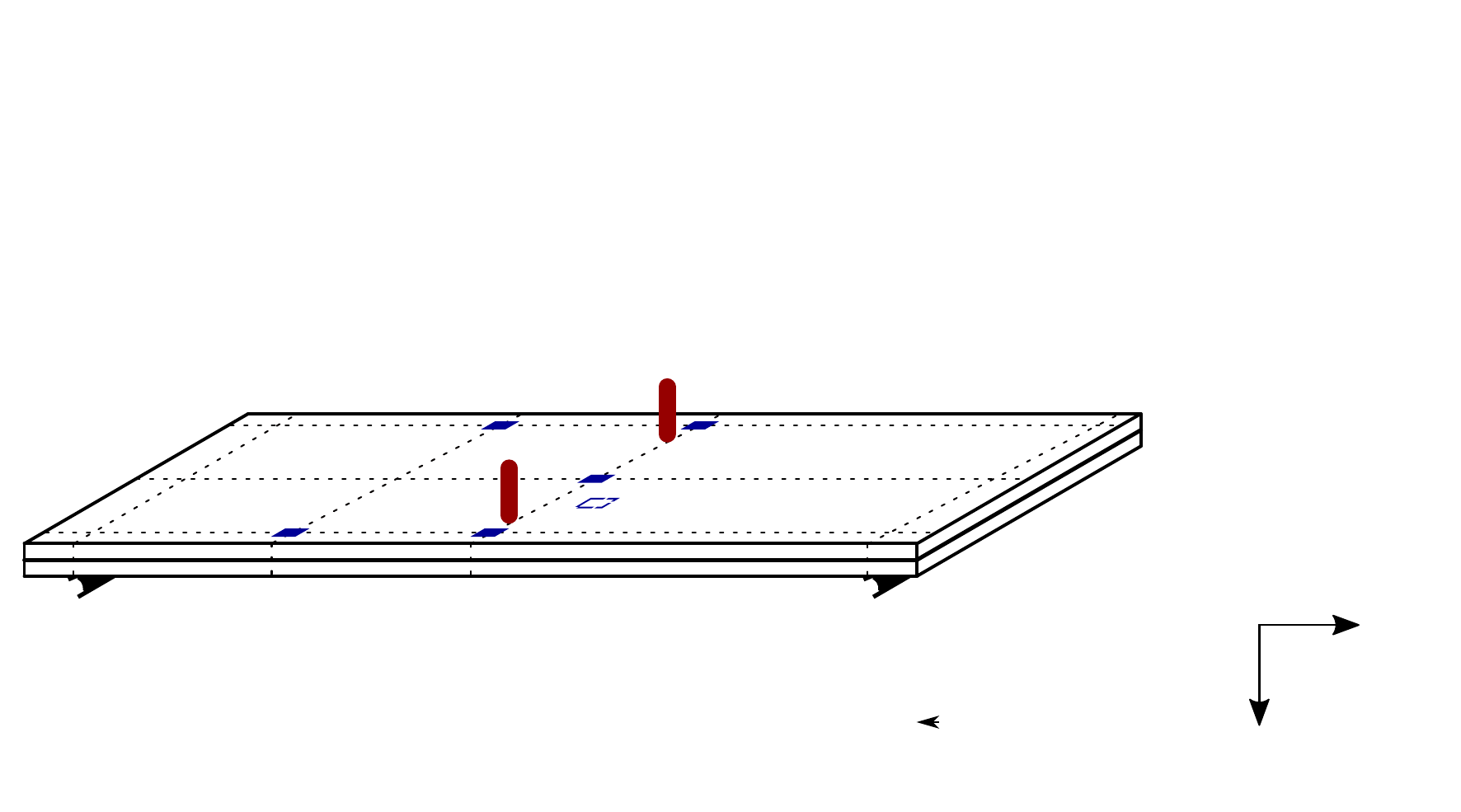}
\\
\\
\includegraphics[width=65mm]{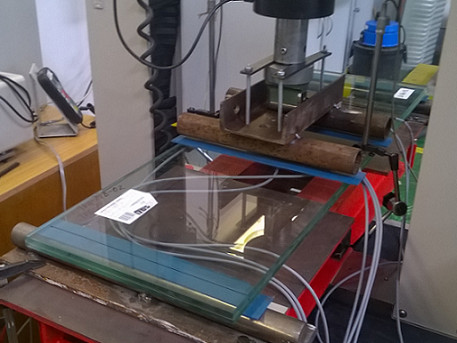}
&
\includegraphics[width=65mm]{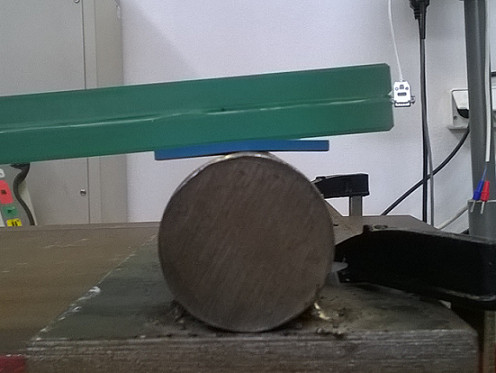} 
\\
\\
\includegraphics[width=65mm]{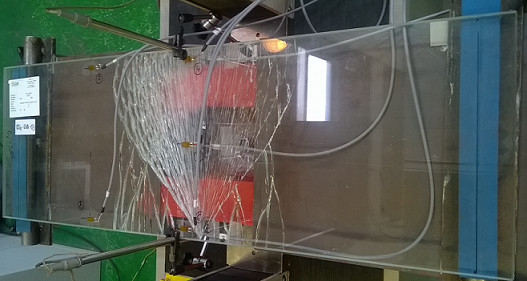} 
&
\includegraphics[width=65mm]{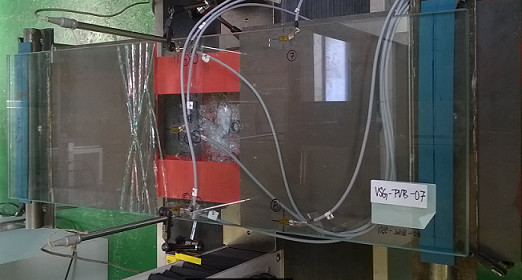} 
\\
\end{tabular}
}
\caption{Experimental setup and crack patterns; courtesy of Tom\'{a}\v{s} H\'{a}na from CTU in Prague.
}
\label{fig:Sphoto2}
\end{figure}

The set of \textit{four-point bending tests} was performed on five ANG-EVA and five ANG-PVB samples at the Faculty of Civil Engineering, Czech Technical University in Prague. The experiments were displacement-controlled with the cross-head speed of the MTS loading device of 0.03 mm/s. The samples were placed on cylindrical supports and separated with rubber pads,~\Fref{fig:Sphoto2}. The measured room temperature was $25\,^{\circ}\mathrm{C}$.
Two displacement sensors measured the vertical deflection at the midspan of samples to check that the experimental set-up was symmetric. Additionally, eight strain gauges LY~11-10/120 were attached to the glass surface: five on the upper surface under compression and three on the bottom surface under tension,~\Fref{fig:Sphoto2} or \cite{hanafour}. 

\Tref{tab:strength} summarizes the extreme tensile failure stresses for the two types of laminated glass samples independently. These values are utilized for the validation of the numerical solver in the next section. Mostly, the fracture originated from a spot on edge.

\begin{table}[!ht]
\centering{
\begin{tabular}{ccc}
\hline
samples & min failure stress [MPa] & max failure stress [MPa] \\
\hline
ANG-EVA & 32 & 60 \\
ANG-PVB & 28 & 69 \\
\hline
\end{tabular}
}
\caption{Extreme tensile failure stresses on the bottom surface of laminated glass under four-point bending.}
\label{tab:strength}
\end{table}

\section{Validation of phase-field model against experimental data} 
\label{S:ESV} 

In order to assess the behaviour of the phase-field model and the quality of the identified material parameters for both foils, we present a validation of the numerical predictions against the experimentally measured response for the three-layer laminated glass plates.

\begin{figure}[ht]
    \centering
    \def\svgwidth{0.9\textwidth}
\begingroup%
  \makeatletter%
  \providecommand\color[2][]{%
    \errmessage{(Inkscape) Color is used for the text in Inkscape, but the package 'color.sty' is not loaded}%
    \renewcommand\color[2][]{}%
  }%
  \providecommand\transparent[1]{%
    \errmessage{(Inkscape) Transparency is used (non-zero) for the text in Inkscape, but the package 'transparent.sty' is not loaded}%
    \renewcommand\transparent[1]{}%
  }%
  \providecommand\rotatebox[2]{#2}%
  \newcommand*\fsize{\dimexpr\f@size pt\relax}%
  \newcommand*\lineheight[1]{\fontsize{\fsize}{#1\fsize}\selectfont}%
  \ifx\svgwidth\undefined%
    \setlength{\unitlength}{333.61772199bp}%
    \ifx\svgscale\undefined%
      \relax%
    \else%
      \setlength{\unitlength}{\unitlength * \real{\svgscale}}%
    \fi%
  \else%
    \setlength{\unitlength}{\svgwidth}%
  \fi%
  \global\let\svgwidth\undefined%
  \global\let\svgscale\undefined%
  \makeatother%
  \begin{picture}(1,0.5611232)%
    \lineheight{1}%
    \setlength\tabcolsep{0pt}%
    \put(0,0){\includegraphics[width=\unitlength,page=1]{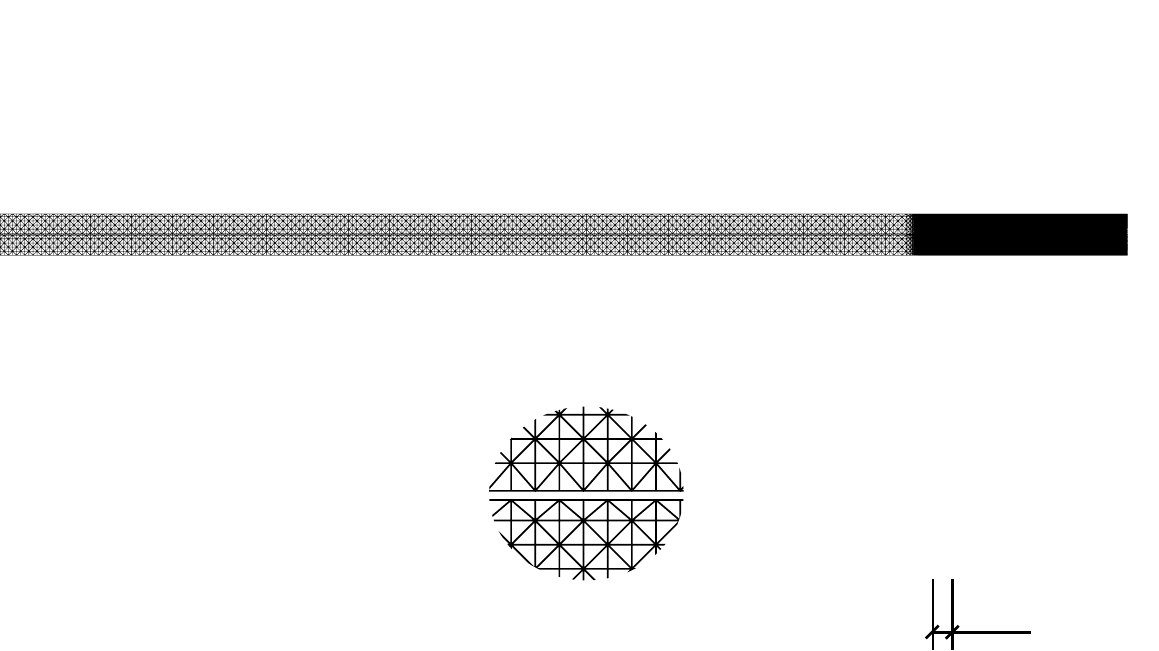}}%
    \put(0.83612419,0.02394936){\color[rgb]{0,0,0}\makebox(0,0)[lt]{\lineheight{1.25}\smash{\begin{tabular}[t]{l}$h=0.25$mm\end{tabular}}}}%
    \put(0.00204198,0.38080403){\color[rgb]{0,0,0}\makebox(0,0)[lt]{\lineheight{1.25}\smash{\begin{tabular}[t]{l}Mesh for $\mathbf{u}$ \end{tabular}}}}%
    \put(0.0019547,0.30399448){\color[rgb]{0,0,0}\makebox(0,0)[lt]{\lineheight{1.25}\smash{\begin{tabular}[t]{l}Mesh for $d$ \end{tabular}}}}%
    \put(0,0){\includegraphics[width=\unitlength,page=2]{figure14.pdf}}%
  \end{picture}%
\endgroup%

    \caption{Locally refined discretization for the 2D laminated glass plane-stress model. Phase-field variable $d$ is calculated on the sub-mesh in both glass layers only.}
    \label{fig:mesh_lg}
\end{figure}

Even though the beam formulation is computationally most effective, we selected for the simulation
the 2D plane-stress model representing the longitudinal cross-section, {see the used mesh in \Fref{fig:mesh_lg}}. The reasons is that this model takes into account the 
transverse compression of the interlayer, and 
provides a better way of
visualisation of the crack evolution and its final pattern. Moreover, the more expensive plate formulation did not provide additional information and improvement as only one crack localised near the loading cylinder, recall~\Sref{S:comp_dim}. 

The phase-field sub-problem, i.e. the damage evolution equation, is defined for the glass layers only {(recall \Fref{fig:mesh_lg})}. Therefore, the phase field is not continuous in the domain, and no crack can evolve in the interlayer. This assumption is in agreement with the displacement-controlled tests when the fractured glass did not break into pieces, as the shards were still connected to the foil, see~\Fref{fig:Sphoto2}.

Note that this modelling choice implies the homogeneous Neumann conditions on glass layer boundaries, recall~\Eref{E:ev_eq}. Using the same mesh for $\boldsymbol{u}$ and $\pfv$ variables would imply the homogeneous Dirichlet boundary conditions at the glass-polymer boundary that result in interfacial crack branching, see, e.g., Figures~21 and 22 in~\cite{sargado2018high} and the accompanying discussion.

\subsection{Four-point bending tests on solid laminated glass samples} 

\begin{figure}[!ht]
    \centering
    \small EVA-based laminated glass plates\\
    \vspace{1mm}
    \begin{tabular}{rl}
        \includegraphics{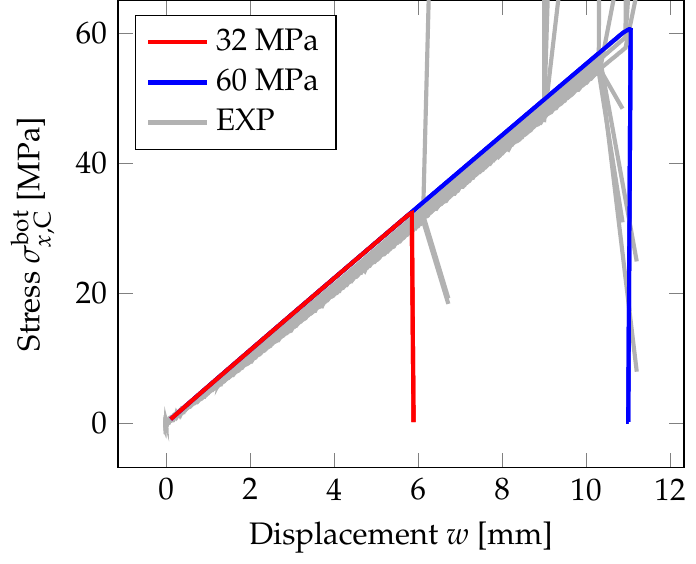}
        &
        \includegraphics{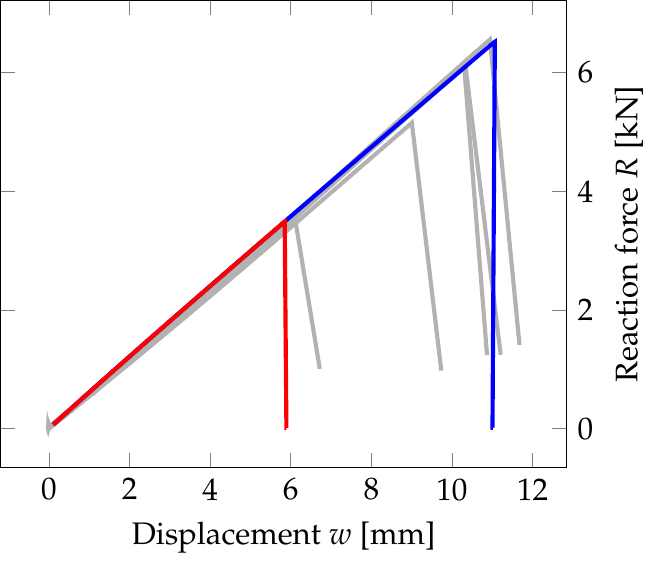}
        \\
        \vspace{0mm}
    \end{tabular}
    \centering
    \small PVB-based laminated glass plates\\
    \vspace{1mm}
    \begin{tabular}{rl}
        \includegraphics{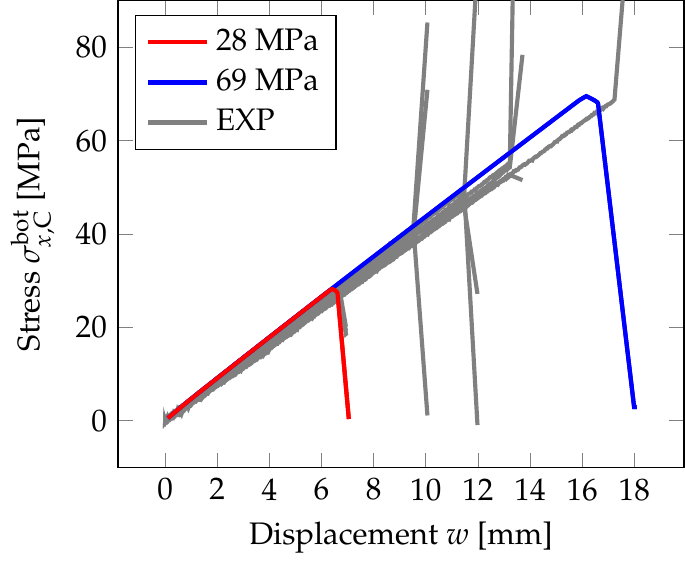}
        &
        \includegraphics{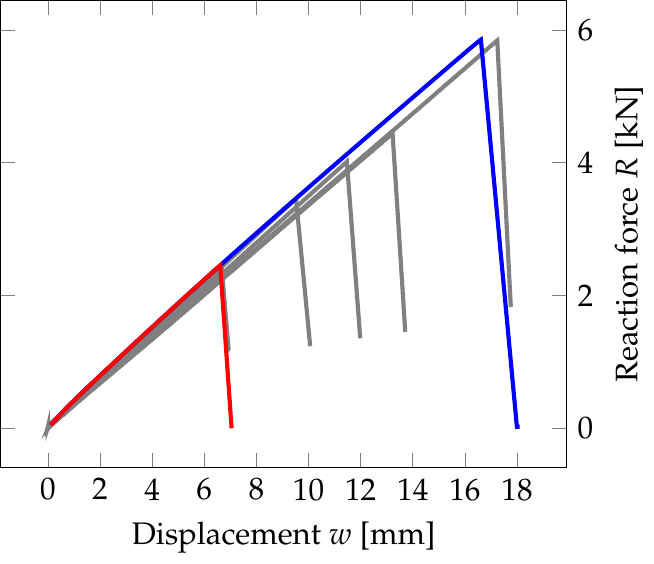}
    \end{tabular}
    \caption{PF-P formulation, anisotropic staggered approach with the spectral-decomposition split, PS~model:
    Evolution of the largest tensile stress at the midpoint and the overall reaction force under the loading points with respect to the mid-point deflection. Experimental data (EXP, grey lines); numerical response for the lowest measured failure stress (red) and for the highest value (blue).}
    \label{F:val}
\end{figure}

The numerical model is validated against the experimental data for the first loading stage, i.e., until the fracture of one glass layer. For laminated glass, the vertical displacements denoted as $w$ in~\Fref{F:val} do not correspond to the prescribed positions of the loading head, but to the deflections at the midspan measured during the bending tests, see~\Fref{fig:Sphoto2}. The fracture energy was set according to~\Eref{eq:lcP} using the extreme tensile strengths measured during the experimental testing, \Tref{tab:strength}. 

For both foils, we achieved a very good agreement with the experimental data for the normal stresses and the overall force reaction of the laminated plate. 
The numerical prediction is slightly stiffer for some of the laminated glass plates. The reasons for this small overestimation could be 
the deviations in material properties or dimensions due to the production tolerances. Because the shear coupling is higher for the numerical model, the error in the failure deflection for the failure stress of 69~MPa is about 4\% and the fracture occurs for slightly lower numerical deflection. On the other hand, the numerical and experimental response corresponding to the minimal failure stress of 28~MPa fits well, the error in deflections is about 1\%. For the EVA-based samples, the error in failure deflections is 1\% for the largest failure stress of 60~MPa and 3\% for the lowest value of 32~MPa.

\begin{figure}[!ht]
\centering
    \begin{tabular}{cc}
     \includegraphics[width=60mm]{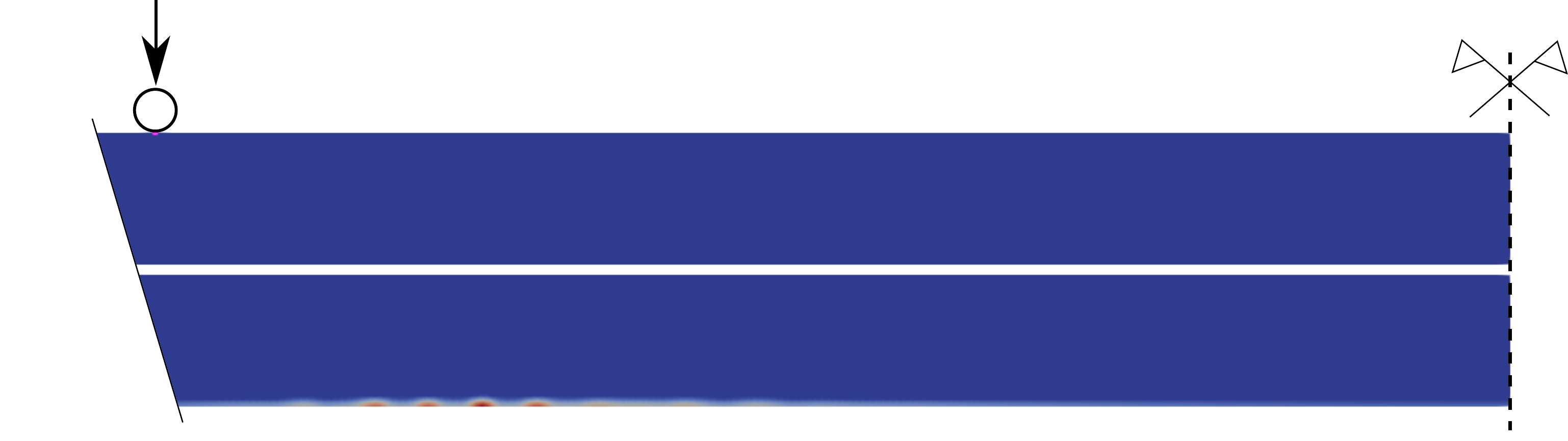} & 
     \includegraphics[width=60mm]{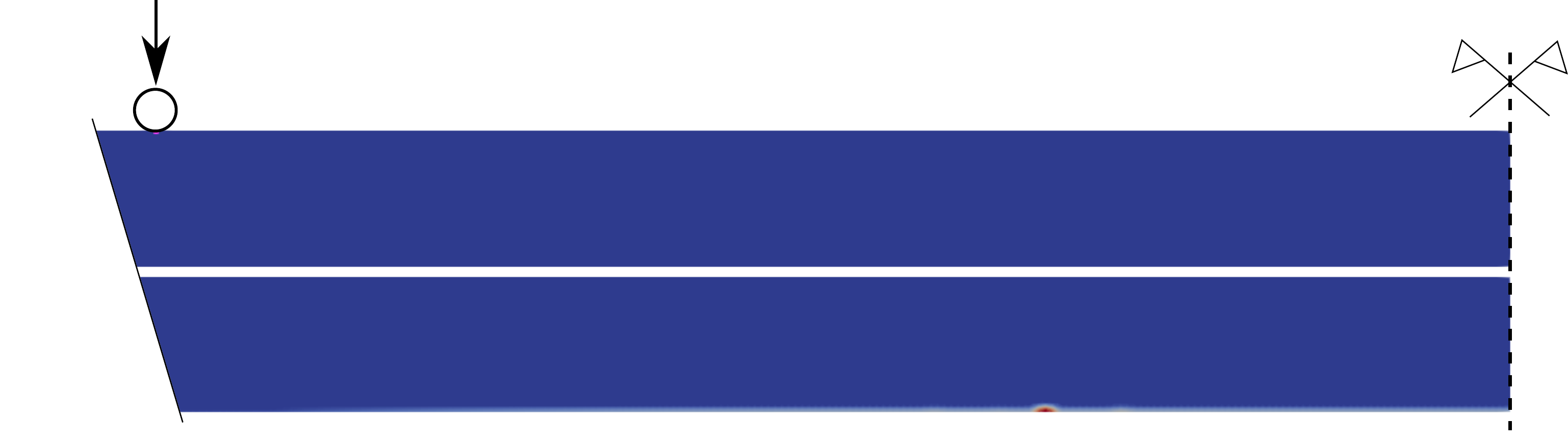} \\
     \includegraphics[width=60mm]{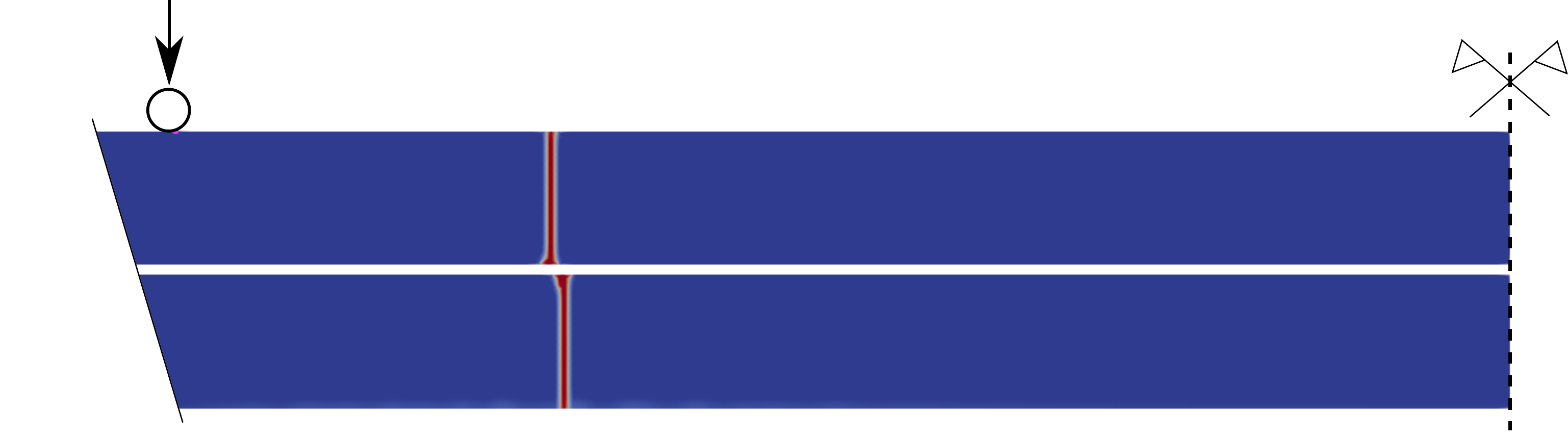} & 
    \includegraphics[width=60mm]{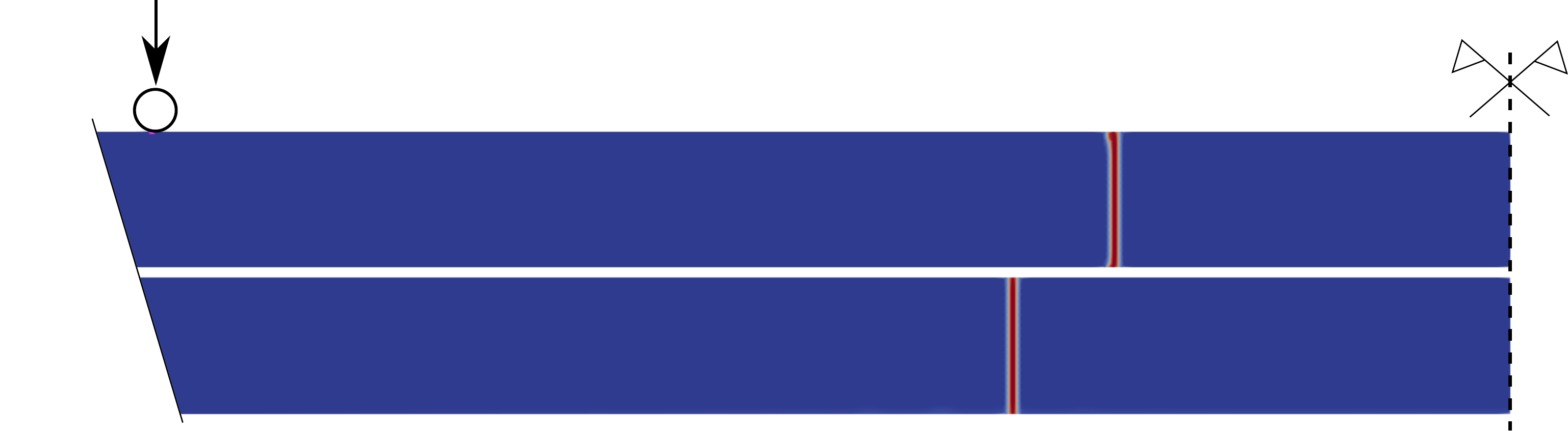}\\
    \small EVA & \small PVB  
    \end{tabular}
    \caption{PF-P formulation, anisotropic staggered approach with the spectral-decomposition split, PS~model: Crack initiation and localised cracks laminated glass samples under four-point bending.}
    \label{fig:3L_dam}
\end{figure}

The numerical prediction of the crack evolution obtained by the phase-field fracture model is shown in~\Fref{fig:3L_dam}. The first row corresponds to the crack initiation and the second to the position of the localised cracks for both types of laminated glass. Even for a very fine increment of the applied displacement $\Delta w = 3 \times 10^{-5}$~mm, the cracks
appear in both glass layers at the same converged step of the staggered algorithm. For EVA-based samples, the phase-field parameter evolves in a few bumps closer to the loading cylinder, and the cracks localised in glass layers are almost above each other. For the PVB-laminated glass, the bottom crack starts to evolve from only one initial point, and the upper crack is shifted to the centre. 

\subsection{Four-point bending tests on laminated glass samples with one layer fractured} 

During the experimental testing, the bottom glass layer was damaged first for the majority of tested samples, and multiple cracks evolved, see~\Fref{fig:fract_1st}. 
Then, the sample was unloaded, and the fracture of the second ply and so the collapse of the laminated glass sample occurred during the second loading stage. 

\begin{figure}[!ht]
\centering
    \begin{tabular}{cc}
    \includegraphics[width=60mm]{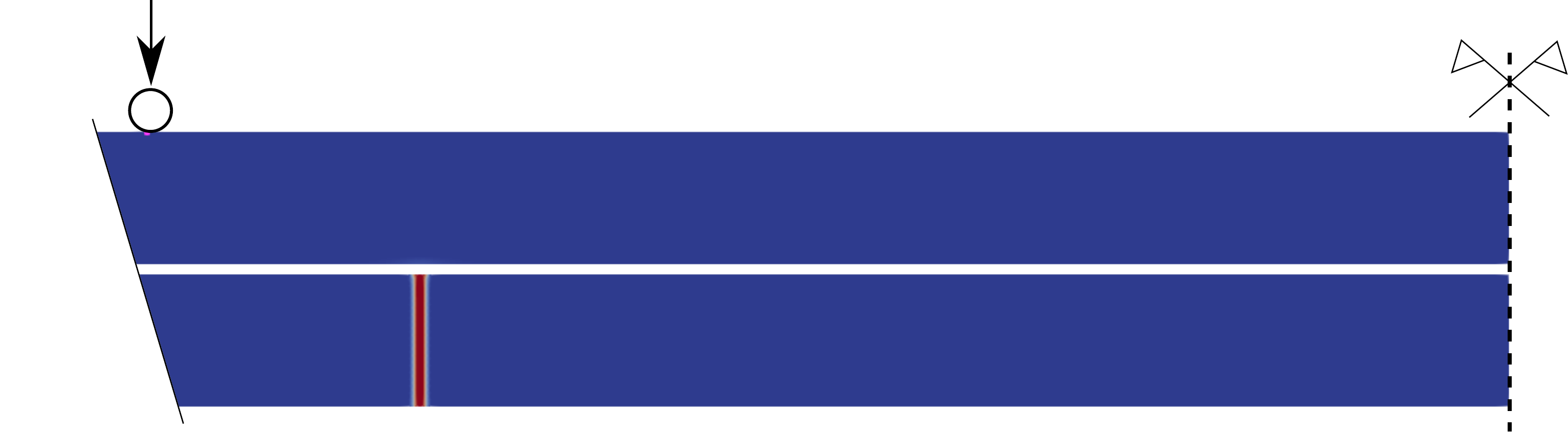} & 
    \includegraphics[width=60mm]{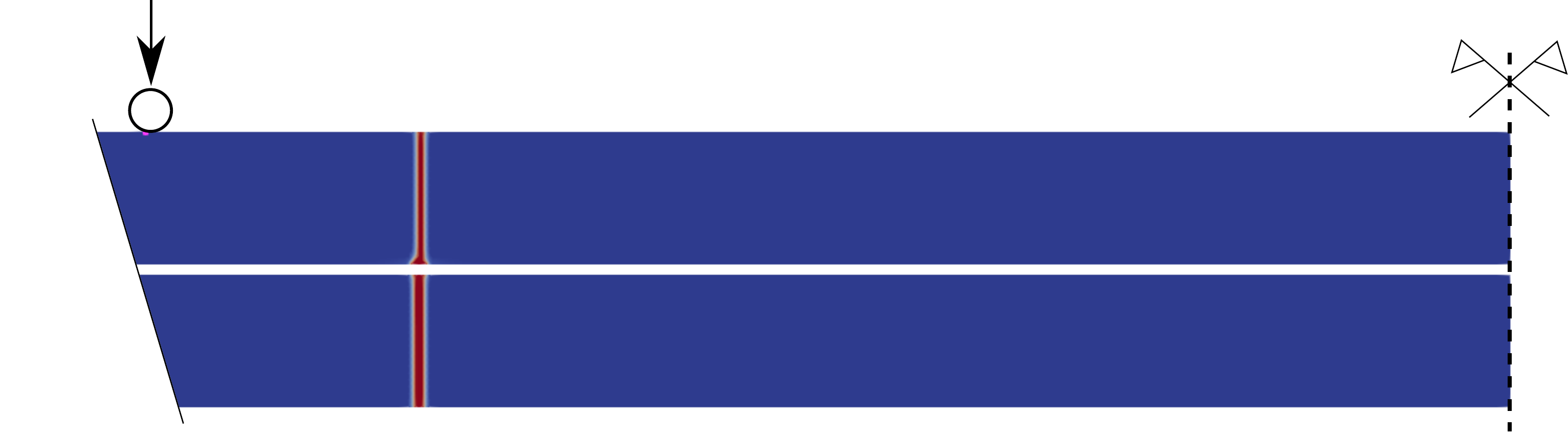} \\
    \includegraphics[width=60mm]{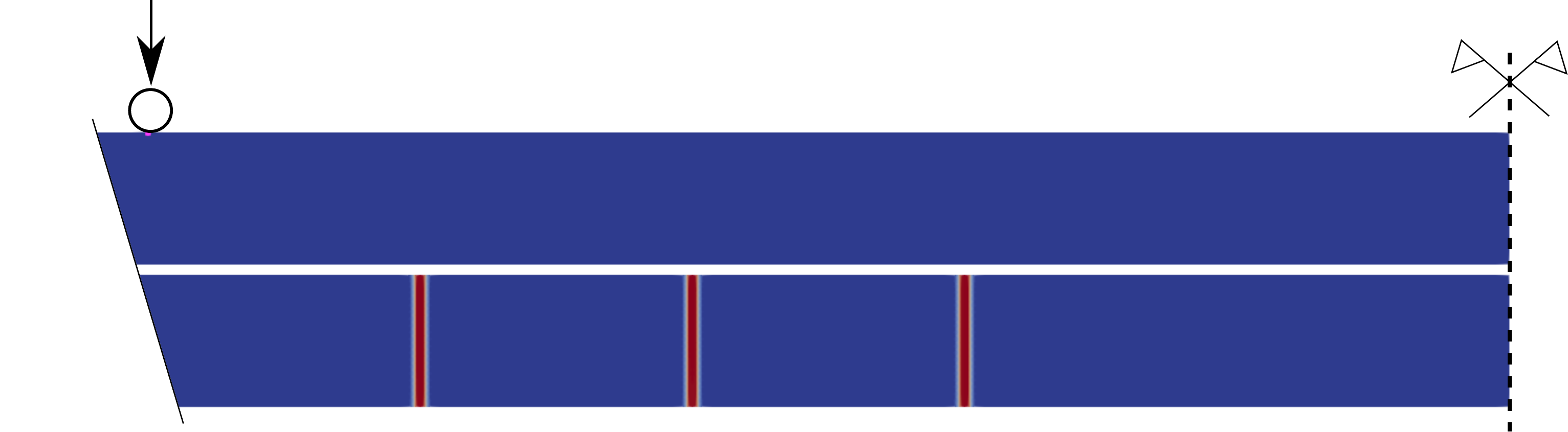} & 
    \includegraphics[width=60mm]{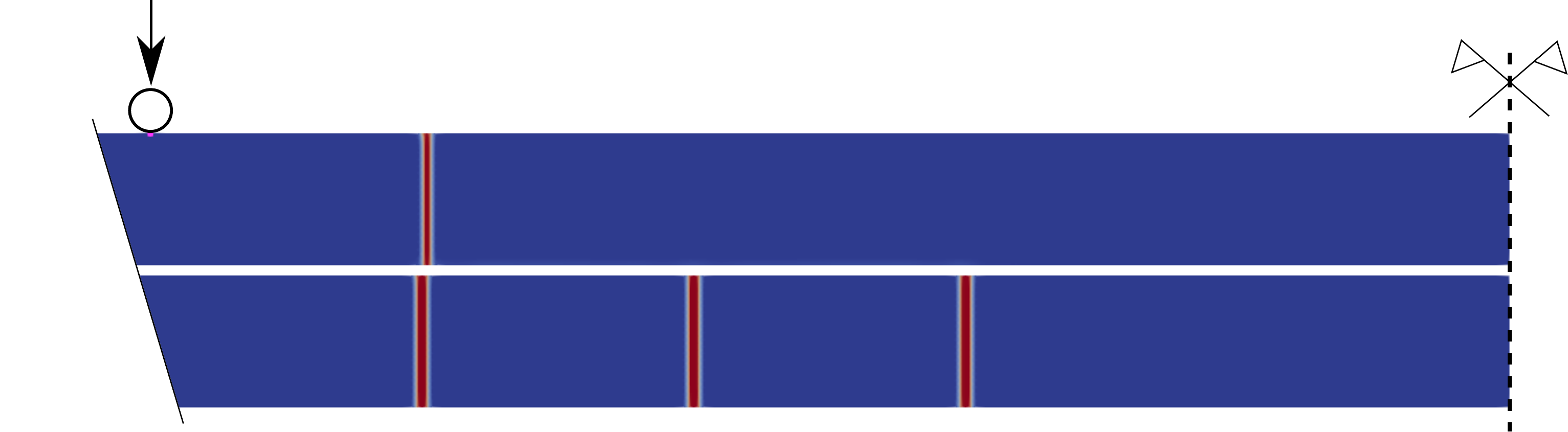} \\
    \includegraphics[width=60mm]{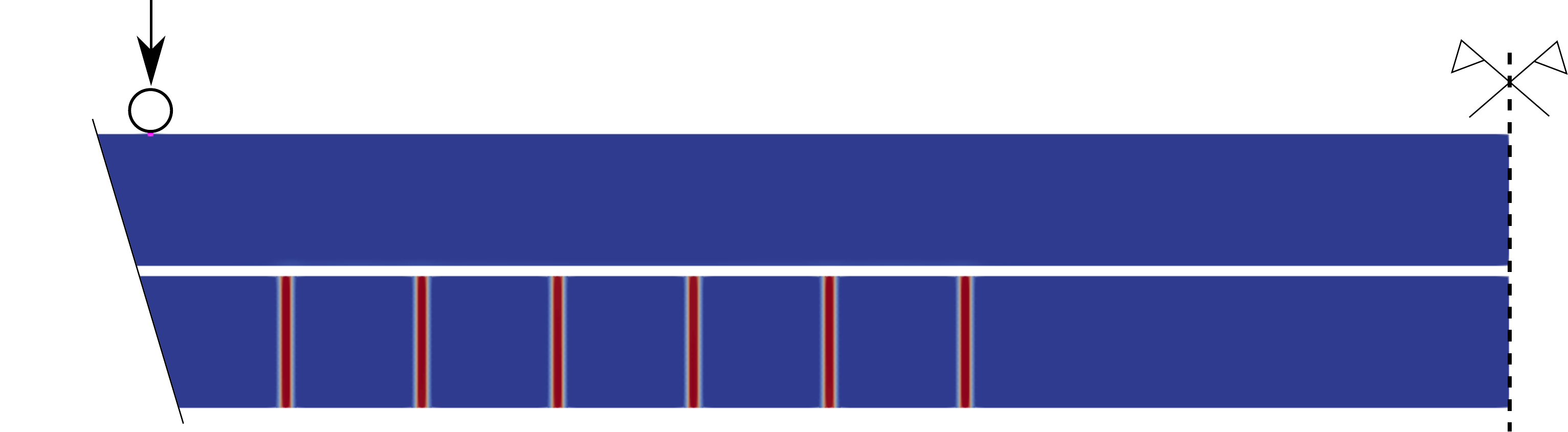} & 
    \includegraphics[width=60mm]{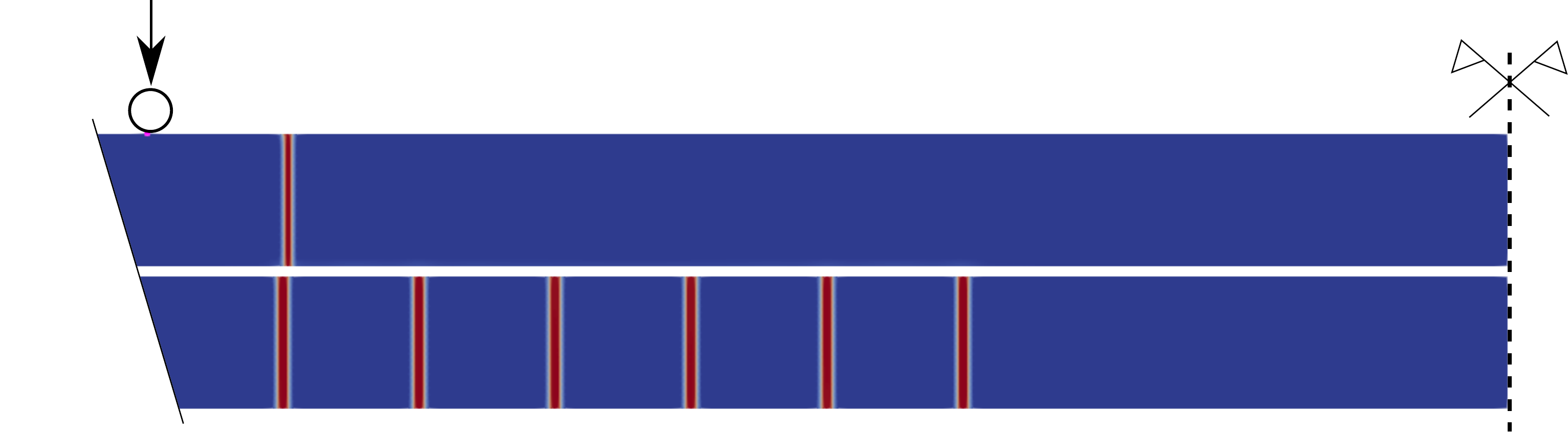} \\
    \small
     initial state & 
     \small
     after localisation  \\
    \end{tabular}
    \caption{Phase-field variable in pre-cracked laminated glass under four-point bending with a different number of initial cracks in a symmetric half of the sample. Initial crack pattern in bottom glass ply and the final fracture for the EVA-based samples for PF-P formulation, anisotropic staggered approach with the spectral-decomposition split, PS~model.}
    \label{fig:3L_dam2}
\end{figure}

\begin{figure}[p]
    \centering
        \begin{tabular}{cccc}
            \rotatebox[origin=c]{90}{\small ANG-EVA-1} 
            &
            \raisebox{-0.5\height}{\includegraphics[width=0.4\textwidth]{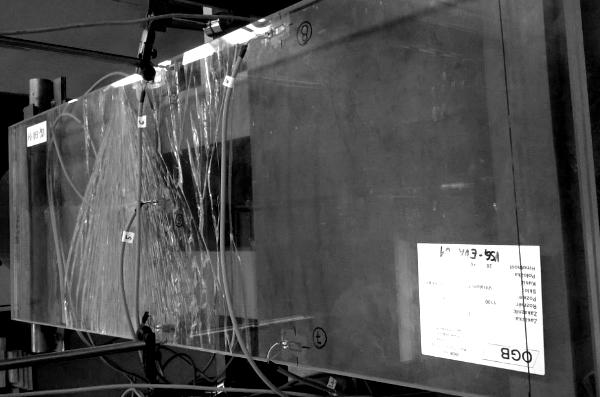}}
            &
            \rotatebox[origin=c]{90}{\small ANG-PVB-1} 
            &
            \raisebox{-0.5\height}{\includegraphics[width=0.4\textwidth]{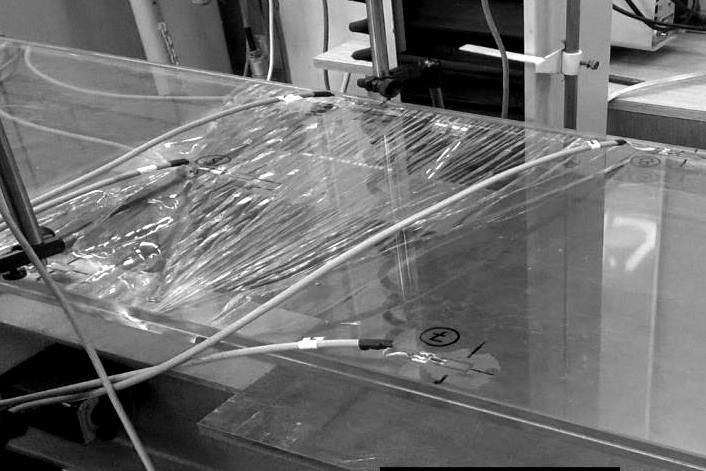}}
            \\
            \\
            \rotatebox[origin=c]{90}{\small ANG-EVA-2} 
            &
            \raisebox{-0.5\height}{\includegraphics[width=0.4\textwidth]{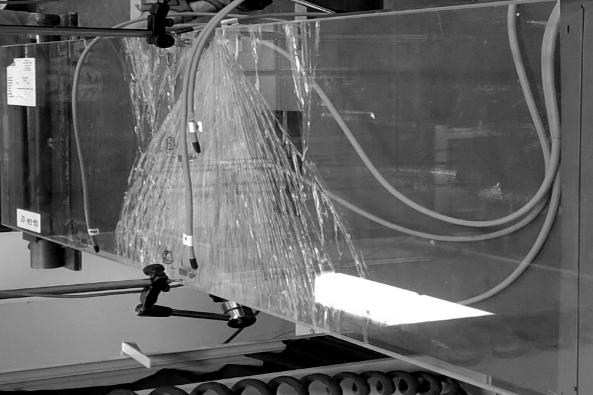}}
            &
            \rotatebox[origin=c]{90}{\small ANG-PVB-2} 
            &
            \raisebox{-0.5\height}{\includegraphics[width=0.4\textwidth]{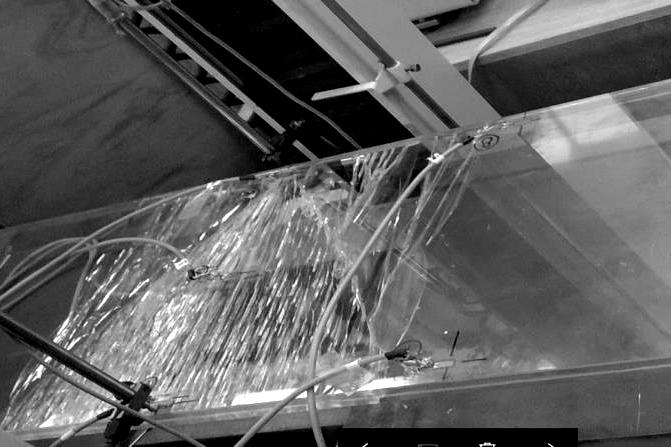}}
            \\
            \\
            \rotatebox[origin=c]{90}{\small ANG-EVA-3} 
            &
            \raisebox{-0.5\height}{\includegraphics[width=0.4\textwidth]{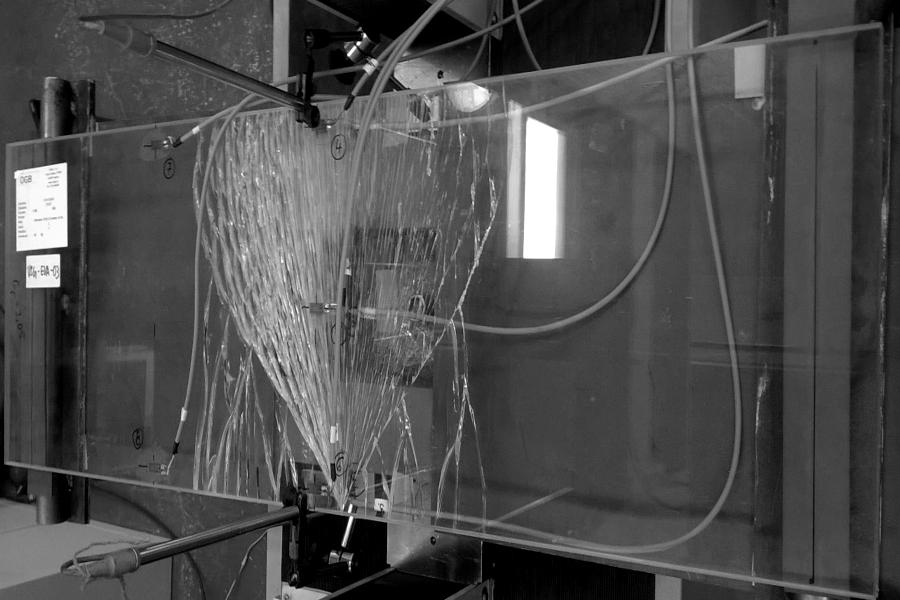}}
            &
            \rotatebox[origin=c]{90}{\small ANG-PVB-3} 
            &
            \raisebox{-0.5\height}{\includegraphics[width=0.4\textwidth]{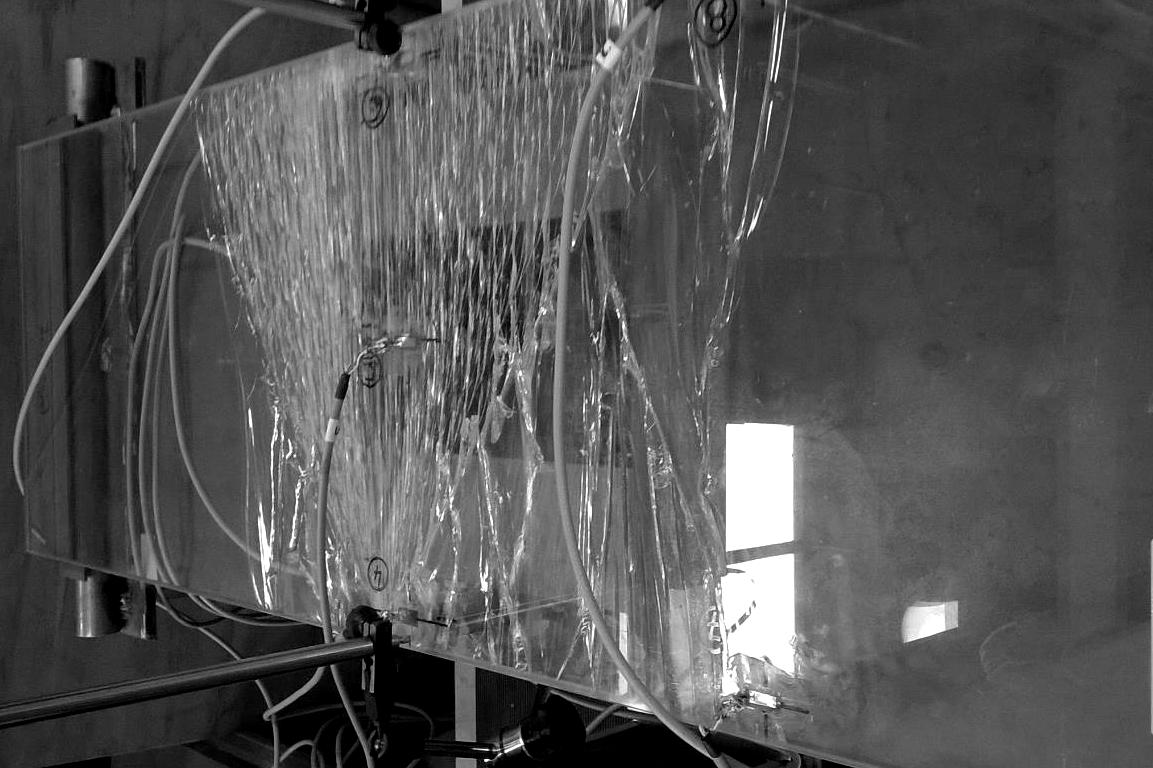}}
            \\
            \\
            \rotatebox[origin=c]{90}{\small ANG-EVA-4} 
            &
            \raisebox{-0.5\height}{\includegraphics[width=0.4\textwidth]{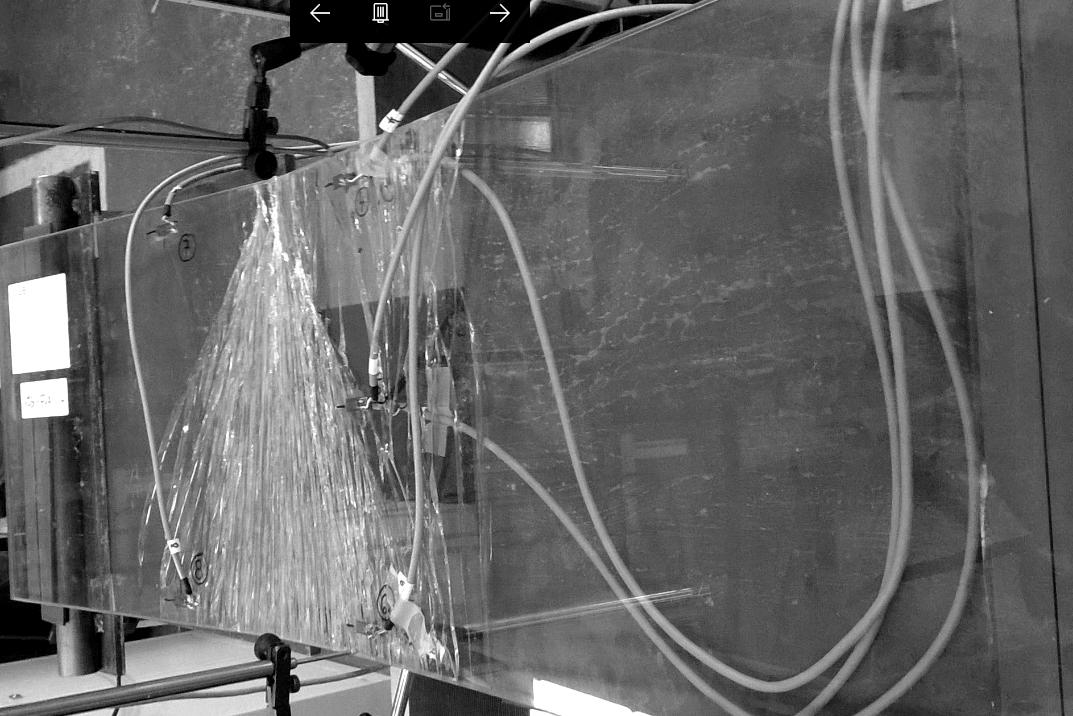}}
            &
            \rotatebox[origin=c]{90}{\small ANG-PVB-4} 
            &
            \raisebox{-0.5\height}{\includegraphics[width=0.4\textwidth]{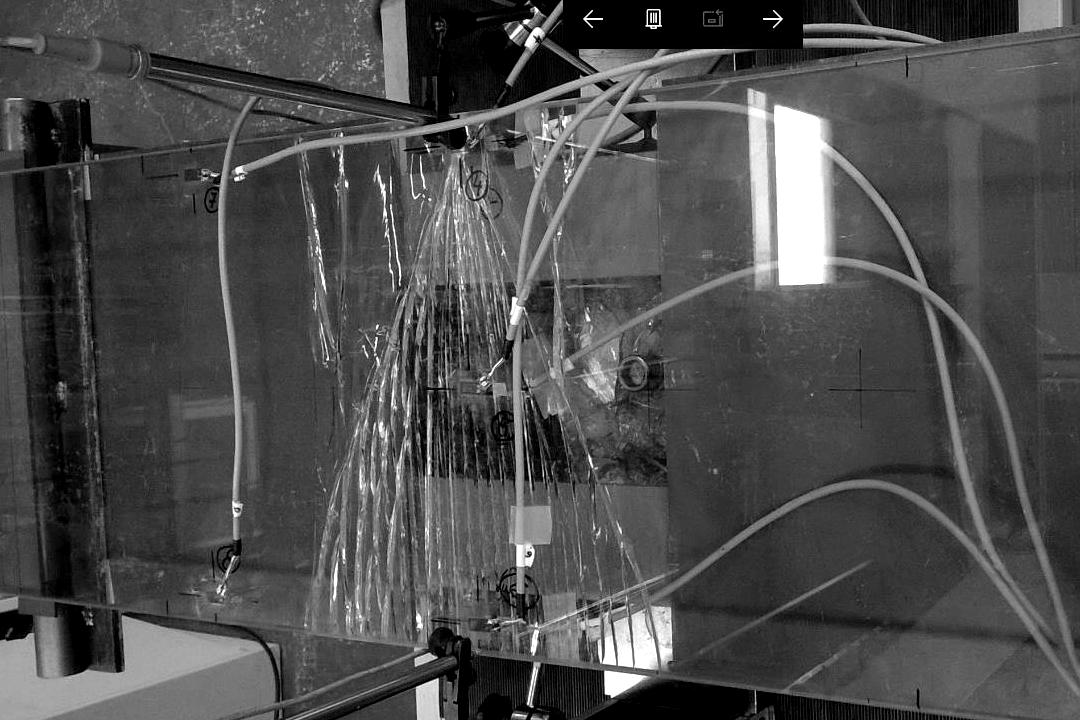}}
            \\
            \\
            \rotatebox[origin=c]{90}{\small ANG-EVA-5} 
            &
            \raisebox{-0.5\height}{\includegraphics[width=0.4\textwidth]{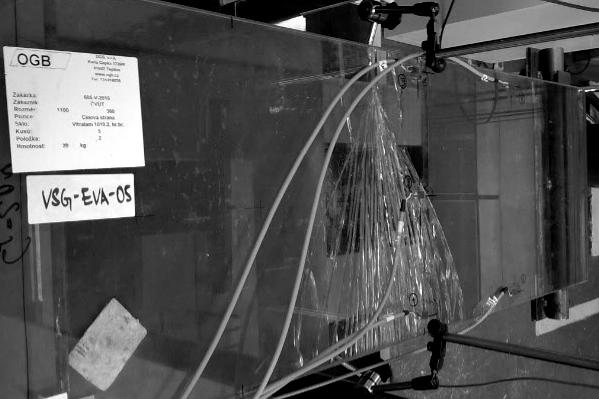}}
            &
            \rotatebox[origin=c]{90}{\small ANG-PVB-5} 
            &
            \raisebox{-0.5\height}{\includegraphics[width=0.4\textwidth]{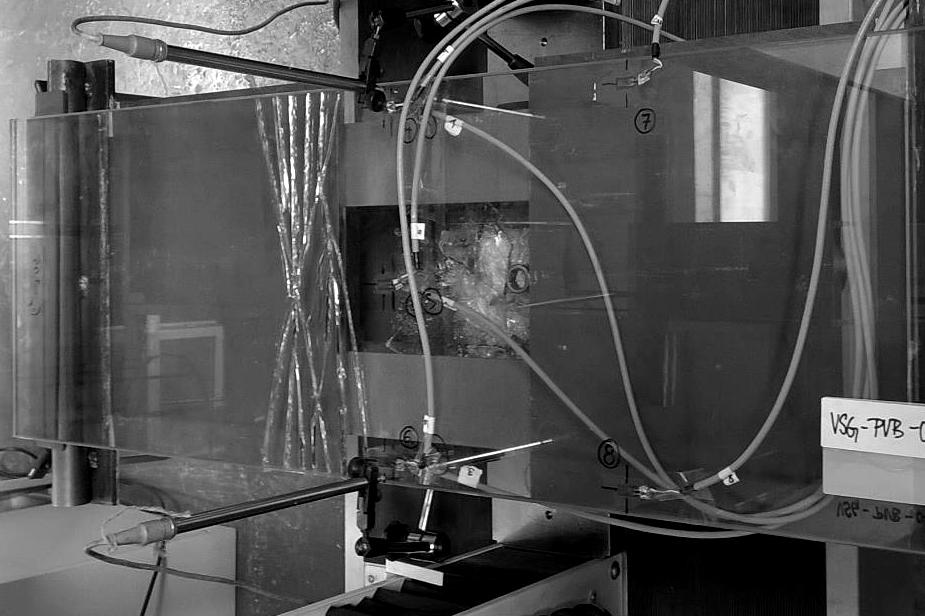}}
        \end{tabular}
    \caption{Fracture pattern after first loading for laminated glass samples. Both glass layers were damaged for ANG-EVA-4, whereas the fracture patterns for the other samples correspond only to the fracture of the bottom glass layer; courtesy of Tom\'{a}\v{s} H\'{a}na from CTU in Prague.}
    \label{fig:fract_1st}
\end{figure}

To numerically simulate this second stage of loading and so the residual resistance of the fractured plate, we defined initial crack patterns in the bottom glass layer, consisting of one, three, or six cracks for a half of the sample, \Fref{fig:3L_dam2}. The cracks were defined by setting the initial phase-field variable to one with the width of the initial cracks $2\lc$. This simulations were performed only for the EVA-based laminated glass samples, and the final crack patterns are shown in~\Fref{fig:3L_dam2}. Again only one crack localised in the upper layer.

\begin{figure}[ht]
    \centering
    \includegraphics{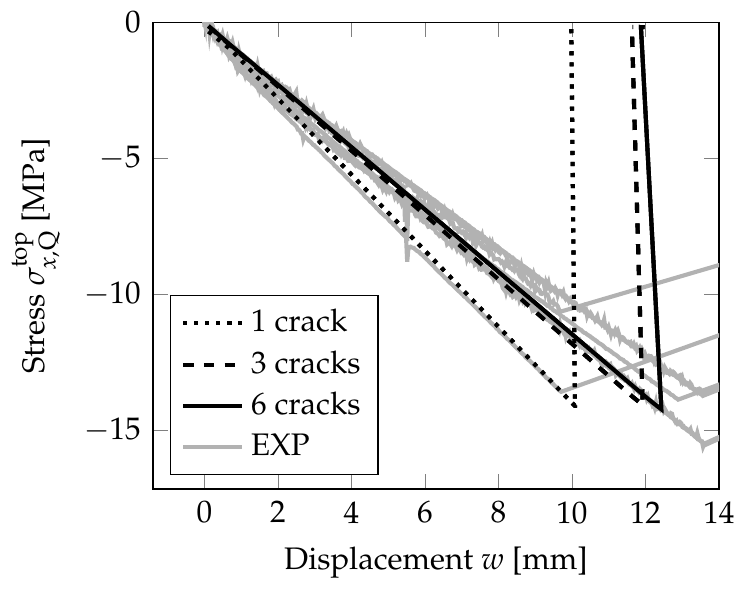}
    \includegraphics{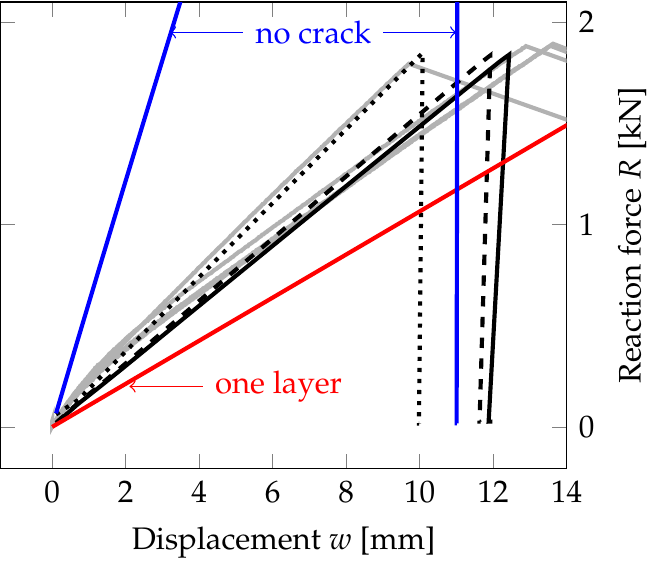}
    \caption{PF-P formulation, anisotropic staggered approach with the spectral-decomposition split, PS~model: Evolution of the compressive stress at a quarter of midspan and the overall reaction under the loading points for the mid-span displacement for the second stage of loading and the EVA-based laminated glass plates. Experimental data (EXP, grey lines) and numerical response for one failure stress of 60~MPa (black) with different number of initial cracks.}
    \label{fig:3L_plots}
\end{figure}

\Fref{fig:3L_plots} illustrates the response of the EVA-laminated glass samples with the bottom layer fractured assuming the largest failure stress of 60~MPa. The evolution of compressive stresses on the upper surface at a quarter of the midspan and the overall reaction are plotted for the mid-span displacement. In this case, the tensile stresses are not validated due to the fracture in the bottom glass and the inaccessible bottom surface of the upper glass layer due to the lamination. 

The force-displacement diagram in~\Fref{fig:3L_plots} shows the residual resistance of the laminated glass plate with one glass damaged.
Two limits bound the stiffness of the samples: the upper limit corresponds to the behaviour of an undamaged laminated glass and the lower bound to the response of a single glass layer.
The numerical model with one initial crack in a symmetric half of the bottom glass is stiffer than most of the experimentally measured responses. Therefore, the critical stress is also reached for a lower prescribed displacement. The force-displacement diagram fits well the experimental response for ANG-EVA-5 in~\Fref{fig:fract_1st}, where the damage after the first stage of loading is the lowest one for the EVA-based samples, and the sample is stiffer due to the triangular shape of the fracture pattern.  
Considering more initial cracks in the bottom glass under the largest bending moment results in the slopes of the stress-displacement and force-displacement diagram that match better the experimental data. The difference in the responses of a sample with three or six cracks (on a half of the area under the largest bending) is small. 
Because we neglected the weight of the fractured glass sample, the numerical post-fracture response does not correspond exactly to that what was observed in the experiment.  

This analysis revealed that the stiffness of the partially fractured laminated glass can be approximated even with a 2D plane-stress model with initially predefined cracks. The numerical model matched the experimentally measured response very well and provides better estimation than a one-glass-layer limit.

\subsection{Beam model for laminated glass and influence of interlayer}

\begin{figure}[ht]
    \centering
    \begin{tabular}{rl}
        \includegraphics{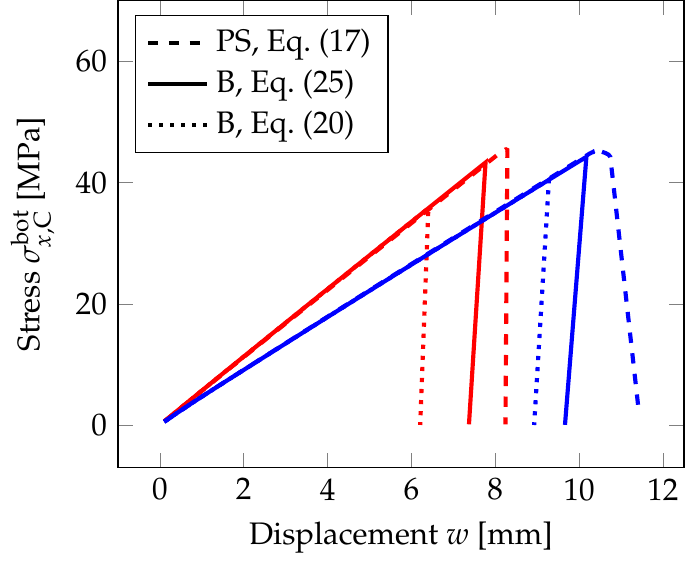}
        &
        \includegraphics{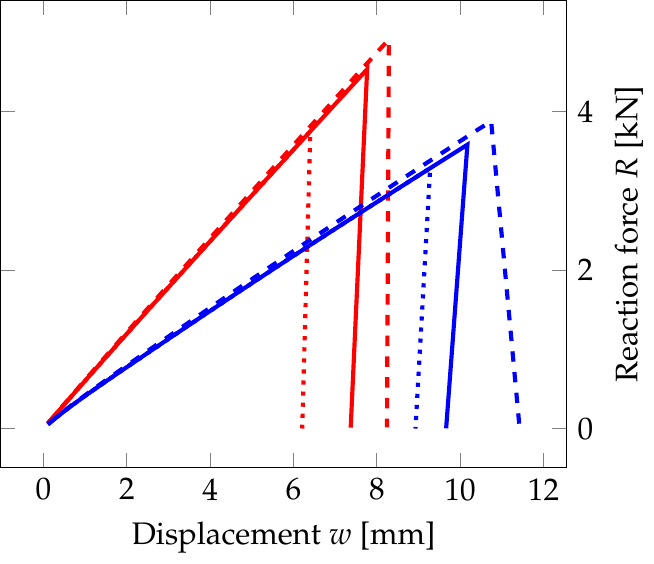}
    \end{tabular}
    \caption{PF-P formulation, anisotropic staggered approach with the spectral-decomposition split:
    Comparison of the response of laminated glass plates with EVA (red line) or PVB (blue line) using the plane-stress (PS) or beam formulation (B). The evolution  of  the  largest  tensile  stress  at the  midpoint  and  the  overall reaction  under  the  loading  points for  the mid-span displacement. The bending strength of glass is set to 45~MPa.}
    \label{F:3L_BvPS}
\end{figure}

Finally, \Fref{F:3L_BvPS} compares the numerical response of EVA-based and PVB-based laminated glass samples using the PS~model and the B~model, corresponding to the plane-stress model of the longitudinal cross-section and to the three-layer beam respectively. For the loading rate of 0.03~mm/s, the EVA interlayer provides better shear coupling, and the response of the laminated glass sample is stiffer than that of the PVB-samples. Therefore, the critical tensile strength is reached earlier; the fracture occurs for lower deflections, but the resistance of the sample is higher for EVA-based samples.

For laminated glass, the relationship in~\Eref{E:lc} binding the fracture energy $\Gf$, the length-scale parameter $\lc$, and the critical stress $\ft$ cannot be applied for the B~model, as it was derived for beams under pure bending only. In reality, normal forces arise from the stress redistribution in both glass layers due to the shear coupling by the interlayer. 
A simple possibility how to overcome this problem is to drive the phase-field evolution in~\Eref{E:ev_eq} by the positive elastic energy density on the surface in tension as the following shows.

As the B-model assumes a constant value of the phase field variable through the thickness, the governing equation of damage evolution \eqref{eq:PFGE} is integrated over the cross section to obtain
\begin{align}
\displaystyle\frac{A}{c_{\alpha}}
\left(
\der{\alpha(\pfv)}{\pfv}
-2 \lc^2 \dod[2]{\pfv(x)}{x} \right) = 
- \frac{\lc}{\Gf}
\der{g(\pfv)}{\pfv}
\int_A\psi^+_\mathrm{e}\,\mathrm{d}A,
\label{eq:goveq_beam}
\end{align}
where $A$ is the transverse cross-sectional area. Therefore, the phase field evolution is driven by the stored energy density integrated over the cross-sectional area under tension instead of the extreme value at the surface fibers. To modify this behaviour, we replace the right-hand side integral with     
\begin{align}
    \int_A\psi^+_\mathrm{e}\,\mathrm{d}A \approx \frac{EA}{2}\mathrm{max}\left(\langle\varepsilon(x, h/2)\rangle^2, \langle\varepsilon(x, -h/2)\rangle^2\right),
    \label{eq:mod_energy}
\end{align}
where the strain $\varepsilon(x,z)=\od{u}{x} + z \od{\varphi}{x}$ is obtained from the centerline horizontal displacement $u$ and cross-sectional rotation $\varphi$. This modification of driving force is performed for each glass layer separately, so a crack evolves through the layer once a critical value is reached on one of its surfaces.  

The comparison in~\Fref{F:3L_BvPS} shows that the through-thickness compression of the interlayer does not effect significantly the response. The differences in the slopes of the force-displacement diagram for both formulations are negligible. The glass fracture occurs for the beam formulation earlier as the criterion is set according to the bottom surface, 
but the differences are considerable smaller than for the parametric studies performed in~\Sref{S:NCS}. 

\section{Conclusions}
\label{S:CON}

Using the phase-field fracture models, we studied the brittle response of monolithic and laminated glass plates under bending with the goal
\begin{itemize}
    \item to compare three different modelling strategies based on the phase-field fracture formulations and analysed the extent of the
    nonlinear part of
    the expected linear pre-fracture response,
    \item to discuss the possible dimensional reduction of the problem together with the setting of the length-scale parameter and adjusted fracture energy of glass,
    \item to illustrate how the fracture for the four-point bending tests initiated from glass edges could be enforced,
    \item to predict the response of laminated glass with one layer fractured.
\end{itemize}
The main contributions of the presented study are:
\begin{itemize}
    \item For thin plates under bending, the 
    nonlinear response 
    due to the damage
    before the crack localisation can be significant for some phase-field formulations. This effect can lead to a considerable overestimation of the response.
    \item 
    The fracture energy of soda-lime-silica  glass results in a  very  small  length-scale  parameter.
    For numerical purposes, the fracture energy should be determined concerning the applied dimensional reduction, loading type, and the value of the length-scale parameter compared with the thickness of the structure under bending. Then, a consistent value of the tensile strength corresponds to the glass fracture.
    \item 
    {Two possible ways of the fracture energy scaling were introduced in this study for beam or plate models. For a one-layer monolith under bending, the scaling relations 
    were derived from the evolution equation using a spatially homogeneous solutions. For multi-layer plates, the phase-field evolution can be driven by a positive energy density on a surface in tension as suggested in this study.} 
    \item  For the plate model, the strength has to be reduced in a band or area of sufficient dimensions and not only at nodes directly on edges to reproduce cracking from edges.
    \item The comparison showed that the numerical model provides a very good agreement with the measured stresses and resistance of laminated glass, even though only one/two cracks localised for all discretisations using the quasi-static solver, whereas multiple cracks evolved during the experiment.  
    \item The stiffness and resistance of the partially fractured laminated glass can be approximated with a 2D plane-stress model with initially predefined cracks. The model matched the experiment very well and provided much better estimation than a one glass layer limit.  
    \item For quasi-static loading of laminated glass, the presented examples also validated the time/temperature-dependent material properties of the Ethylene-Vinyl Acetate or PolyVinyl Butyral interlayers derived recently by the authors in~\cite{hana2019experimental}.
\end{itemize}

The next step of this study will be to extend the model to incorporate the effect of the stochastic strength of glass.

\vspace{6pt} 



\authorcontributions{
Conceptualization, A.Z. and J.S.; methodology, A.Z., J.S. and J.Z.; software, J.S.; validation, J.S.; formal analysis, A.Z. and M.Š.; investigation, J.S., A.Z. and J.Z.; writing–original draft preparation, A.Z. and J.S.; writing–review and editing, M.Š., J.Z.; visualization, J.S. and A.Z.; supervision, M.Š. and J.Z..; project administration, M.Š.
}

\funding{
This research was funded by the Czech Science Foundation, the grant number 19-15326S. JZ also acknowledges the partial support by the European Regional Development Fund through the Center of Advanced Applied Sciences at the Czech Technical University in Prague (project No. CZ.02.1.01/0.0/0.0/16\_019/0000778).}

\acknowledgments{
Authors are grateful to Tom\'{a}\v{s} H\'{a}na, Martina Eli\'{a}\v{s}ov\'{a}, and Zden\v{e}k Sokol from the Czech Technical University in Prague for performing the experimental tests reported in this paper. The authors would also like to acknowledge J\'{e}r\'{e}my Bleyer from the  Navier Laboratory, University of Paris-Est for his valuable advice on the FEniCS implementation.
}

\conflictsofinterest{The authors declare no conflict of interest. 
} 

\reftitle{References}


\begin{thebibliography}{-------}
\providecommand{\natexlab}[1]{#1}

\bibitem[Haldimann \em{et~al.}(2008)Haldimann, Luible, and
  Overend]{Haldimann:2008:SUG}
Haldimann, M.; Luible, A.; Overend, M.
\newblock {\em Structural Use of Glass}; Vol.~10, {\em Structural Engineering
  Documents}, {IABSE}: Z\"{u}rich, Switzerland,  2008.

\bibitem[Ledbetter \em{et~al.}(2006)Ledbetter, Walker, and
  Keiller]{ledbetter2006structural}
Ledbetter, S.R.; Walker, A.R.; Keiller, A.P.
\newblock Structural use of glass.
\newblock {\em Journal of Architectural Engineering} {\bf 2006}, {\em
  12},~137--149.

\bibitem[Zhao \em{et~al.}(2019)Zhao, Yang, Wang, and
  Azim]{zhao2019experimental}
Zhao, C.; Yang, J.; Wang, X.e.; Azim, I.
\newblock Experimental investigation into the post-breakage performance of
  pre-cracked laminated glass plates.
\newblock {\em Construction and Building Materials} {\bf 2019}, {\em
  224},~996--1006.

\bibitem[Bonati \em{et~al.}(2019)Bonati, Pisano, and
  Carfagni]{bonati2019redundancy}
Bonati, A.; Pisano, G.; Carfagni, G.R.
\newblock Redundancy and robustness of brittle laminated plates. Overlooked
  aspects in structural glass.
\newblock {\em Composite Structures} {\bf 2019}, {\em 227},~111288.

\bibitem[Overend \em{et~al.}(2007)Overend, De~Gaetano, and
  Haldimann]{overend2007diagnostic}
Overend, M.; De~Gaetano, S.; Haldimann, M.
\newblock Diagnostic interpretation of glass failure.
\newblock {\em Structural Engineering International} {\bf 2007}, {\em
  17},~151--158.

\bibitem[Teotia and Soni(2018)]{teotia2018applications}
Teotia, M.; Soni, R.
\newblock Applications of finite element modelling in failure analysis of
  laminated glass composites: A review.
\newblock {\em Engineering Failure Analysis} {\bf 2018}, {\em 94},~412--437.

\bibitem[Chen \em{et~al.}(2017)Chen, Zang, Wang, Yoshimura, and
  Yamada]{chen2017numerical}
Chen, S.; Zang, M.; Wang, D.; Yoshimura, S.; Yamada, T.
\newblock Numerical analysis of impact failure of automotive laminated glass: a
  review.
\newblock {\em Composites Part B: Engineering} {\bf 2017}, {\em 122},~47--60.

\bibitem[Calderone \em{et~al.}(2009)Calderone, Davies, Bennison, Xiaokun, and
  Gang]{calderone2009effective}
Calderone, I.; Davies, P.; Bennison, S.J.; Xiaokun, H.; Gang, L.
\newblock Effective laminate thickness for the design of laminated glass.
\newblock {\em Glass Performance Days, Tampere, Finland, June} {\bf 2009}, pp.
  12--15.

\bibitem[Galuppi and Royer-Carfagni(2012{\natexlab{a}})]{galuppi2012effective}
Galuppi, L.; Royer-Carfagni, G.F.
\newblock Effective thickness of laminated glass beams: new expression via a
  variational approach.
\newblock {\em Engineering Structures} {\bf 2012}, {\em 38},~53--67.

\bibitem[Galuppi and Royer-Carfagni(2012{\natexlab{b}})]{galuppi2012effectiveP}
Galuppi, L.; Royer-Carfagni, G.
\newblock The effective thickness of laminated glass plates.
\newblock {\em Journal of Mechanics of Materials and Structures} {\bf 2012},
  {\em 7},~375--400.

\bibitem[Galuppi and Royer-Carfagni(2014)]{galuppi2014enhanced}
Galuppi, L.; Royer-Carfagni, G.
\newblock Enhanced effective thickness of multi-layered laminated glass.
\newblock {\em Composites Part B: Engineering} {\bf 2014}, {\em 64},~202--213.

\bibitem[Vallabhan \em{et~al.}(1993)Vallabhan, Das, Magdi, Asik, and
  Bailey]{vallabhan1993analysis}
Vallabhan, C.G.; Das, Y.; Magdi, M.; Asik, M.; Bailey, J.R.
\newblock Analysis of laminated glass units.
\newblock {\em Journal of Structural Engineering} {\bf 1993}, {\em
  119},~1572--1585.

\bibitem[A{\c{s}}{\i}k(2003)]{acsik2003laminated}
A{\c{s}}{\i}k, M.Z.
\newblock Laminated glass plates: revealing of nonlinear behavior.
\newblock {\em Computers \& Structures} {\bf 2003}, {\em 81},~2659--2671.

\bibitem[A{\c{s}}{\i}k and Tezcan(2005)]{acsik2005mathematical}
A{\c{s}}{\i}k, M.Z.; Tezcan, S.
\newblock A mathematical model for the behavior of laminated glass beams.
\newblock {\em Computers \& Structures} {\bf 2005}, {\em 83},~1742--1753.

\bibitem[Ivanov(2006)]{ivanov2006analysis}
Ivanov, I.V.
\newblock Analysis, modelling, and optimization of laminated glasses as plane
  beam.
\newblock {\em International Journal of Solids and Structures} {\bf 2006}, {\em
  43},~6887--6907.

\bibitem[Foraboschi(2007)]{foraboschi2007behavior}
Foraboschi, P.
\newblock Behavior and failure strength of laminated glass beams.
\newblock {\em Journal of Engineering Mechanics} {\bf 2007}, {\em
  133},~1290--1301.

\bibitem[Koutsawa \em{et~al.}(2007)Koutsawa et~al.]{koutsawa2007static}
Koutsawa, Y.; others.
\newblock Static and free vibration analysis of laminated glass beam on
  viscoelastic supports.
\newblock {\em International Journal of Solids and Structures} {\bf 2007}, {\em
  44},~8735--8750.

\bibitem[Foraboschi(2012)]{foraboschi2012analytical}
Foraboschi, P.
\newblock Analytical model for laminated-glass plate.
\newblock {\em Composites Part B: Engineering} {\bf 2012}, {\em
  43},~2094--2106.

\bibitem[Schulze \em{et~al.}(2012)Schulze, Pander, Naumenko, and
  Altenbach]{schulze2012analysis}
Schulze, S.H.; Pander, M.; Naumenko, K.; Altenbach, H.
\newblock Analysis of laminated glass beams for photovoltaic applications.
\newblock {\em International Journal of Solids and Structures} {\bf 2012}, {\em
  49},~2027--2036.

\bibitem[Lancioni and Royer-Carfagni(2009)]{lancioni2009variational}
Lancioni, G.; Royer-Carfagni, G.
\newblock The variational approach to fracture mechanics. {A} practical
  application to the {F}rench {P}anth{\'e}on in {P}aris.
\newblock {\em Journal of Elasticity} {\bf 2009}, {\em 95},~1--30.

\bibitem[Mart{\'\i}nez-Pa{\~n}eda \em{et~al.}(2018)Mart{\'\i}nez-Pa{\~n}eda,
  Golahmar, and Niordson]{martinez2018phase}
Mart{\'\i}nez-Pa{\~n}eda, E.; Golahmar, A.; Niordson, C.F.
\newblock A phase field formulation for hydrogen assisted cracking.
\newblock {\em Computer Methods in Applied Mechanics and Engineering} {\bf
  2018}, {\em 342},~742--761.

\bibitem[Natarajan \em{et~al.}(2019)Natarajan, Annabattula,
  et~al.]{natarajan2019modeling}
Natarajan, S.; Annabattula, R.K.; others.
\newblock Modeling crack propagation in variable stiffness composite laminates
  using the phase field method.
\newblock {\em Composite Structures} {\bf 2019}, {\em 209},~424--433.

\bibitem[Freddi and Mingazzi(2020)]{freddi2020phase}
Freddi, F.; Mingazzi, L.
\newblock Phase Field Simulation of Laminated Glass Beam.
\newblock {\em Materials} {\bf 2020}, {\em 13},~3218.

\bibitem[H{\'a}na \em{et~al.}(2019)H{\'a}na, Janda, Schmidt, Zemanov{\'a},
  {\v{S}}ejnoha, Eli{\'a}{\v{s}}ov{\'a}, and
  Vok{\'a}{\v{c}}]{hana2019experimental}
H{\'a}na, T.; Janda, T.; Schmidt, J.; Zemanov{\'a}, A.; {\v{S}}ejnoha, M.;
  Eli{\'a}{\v{s}}ov{\'a}, M.; Vok{\'a}{\v{c}}, M.
\newblock Experimental and numerical study of viscoelastic properties of
  polymeric interlayers used for laminated glass: Determination of material
  parameters.
\newblock {\em Materials} {\bf 2019}, {\em 12},~2241.

\bibitem[Griffith(1921)]{griffith1921vi}
Griffith, A.A.
\newblock The phenomena of rupture and flow in solids.
\newblock {\em Philosophical Transactions of the Royal Society of London A}
  {\bf 1921}, {\em 221},~163--198.

\bibitem[Bourdin \em{et~al.}(2000)Bourdin, Francfort, and
  Marigo]{bourdin2000numerical}
Bourdin, B.; Francfort, G.A.; Marigo, J.J.
\newblock Numerical experiments in revisited brittle fracture.
\newblock {\em Journal of the Mechanics and Physics of Solids} {\bf 2000}, {\em
  48},~797--826.

\bibitem[Francfort and Marigo(1998)]{francfort1998revisiting}
Francfort, G.A.; Marigo, J.J.
\newblock Revisiting brittle fracture as an energy minimization problem.
\newblock {\em Journal of the Mechanics and Physics of Solids} {\bf 1998}, {\em
  46},~1319--1342.

\bibitem[Amor \em{et~al.}(2009)Amor, Marigo, and Maurini]{amor2009regularized}
Amor, H.; Marigo, J.J.; Maurini, C.
\newblock Regularized formulation of the variational brittle fracture with
  unilateral contact: Numerical experiments.
\newblock {\em Journal of the Mechanics and Physics of Solids} {\bf 2009}, {\em
  57},~1209--1229.

\bibitem[Wu \em{et~al.}(2018)Wu, Nguyen, Nguyen, Sutula, Bordas, and
  Sinaie]{wu2018phase}
Wu, J.Y.; Nguyen, V.P.; Nguyen, C.T.; Sutula, D.; Bordas, S.; Sinaie, S.
\newblock Phase field modeling of fracture.
\newblock {\em Advances in Applied Mechancis: Multi-scale Theory and
  Computation} {\bf 2018}, {\em 52}.

\bibitem[Kiendl \em{et~al.}(2016)Kiendl, Ambati, De~Lorenzis, Gomez, and
  Reali]{kiendl2016phase}
Kiendl, J.; Ambati, M.; De~Lorenzis, L.; Gomez, H.; Reali, A.
\newblock Phase-field description of brittle fracture in plates and shells.
\newblock {\em Computer Methods in Applied Mechanics and Engineering} {\bf
  2016}, {\em 312},~374--394.

\bibitem[Miehe \em{et~al.}(2015)Miehe, Hofacker, Sch{\"a}nzel, and
  Aldakheel]{miehe2015phaseII}
Miehe, C.; Hofacker, M.; Sch{\"a}nzel, L.M.; Aldakheel, F.
\newblock Phase field modeling of fracture in multi-physics problems. Part II.
  Coupled brittle-to-ductile failure criteria and crack propagation in
  thermo-elastic--plastic solids.
\newblock {\em Computer Methods in Applied Mechanics and Engineering} {\bf
  2015}, {\em 294},~486--522.

\bibitem[Bourdin \em{et~al.}(2008)Bourdin, Francfort, and
  Marigo]{bourdin2008variational}
Bourdin, B.; Francfort, G.A.; Marigo, J.J.
\newblock The variational approach to fracture.
\newblock {\em Journal of Elasticity} {\bf 2008}, {\em 91},~5--148.

\bibitem[Miehe \em{et~al.}(2010)Miehe, Welschinger, and
  Hofacker]{miehe2010thermodynamically}
Miehe, C.; Welschinger, F.; Hofacker, M.
\newblock Thermodynamically consistent phase-field models of fracture:
  Variational principles and multi-field FE implementations.
\newblock {\em International Journal for Numerical Methods in Engineering} {\bf
  2010}, {\em 83},~1273--1311.

\bibitem[Miehe \em{et~al.}(2015)Miehe, Sch\"anzel, and Ulmer]{miehe2015phase}
Miehe, C.; Sch\"anzel, L.M.; Ulmer, H.
\newblock Phase field modeling of fracture in multi-physics problems. Part I.
  Balance of crack surface and failure criteria for brittle crack propagation
  in thermo-elastic solids.
\newblock {\em Computer Methods in Applied Mechanics and Engineering} {\bf
  2015}, {\em 294},~449--485.

\bibitem[Pham \em{et~al.}(2011)Pham, Amor, Marigo, and
  Maurini]{pham2011gradient}
Pham, K.; Amor, H.; Marigo, J.J.; Maurini, C.
\newblock Gradient damage models and their use to approximate brittle fracture.
\newblock {\em International Journal of Damage Mechanics} {\bf 2011}, {\em
  20},~618--652.

\bibitem[Mandal \em{et~al.}(2019)Mandal, Nguyen, and Wu]{mandal2019length}
Mandal, T.K.; Nguyen, V.P.; Wu, J.Y.
\newblock Length scale and mesh bias sensitivity of phase-field models for
  brittle and cohesive fracture.
\newblock {\em Engineering Fracture Mechanics} {\bf 2019}, {\em 217},~106532.

\bibitem[Kuhn \em{et~al.}(2015)Kuhn, Schl{\"u}ter, and
  M{\"u}ller]{kuhn2015degradation}
Kuhn, C.; Schl{\"u}ter, A.; M{\"u}ller, R.
\newblock On degradation functions in phase field fracture models.
\newblock {\em Computational Materials Science} {\bf 2015}, {\em
  108},~374--384.

\bibitem[Zhang \em{et~al.}(2017)Zhang, Vignes, Sloan, and
  Sheng]{zhang2017numerical}
Zhang, X.; Vignes, C.; Sloan, S.W.; Sheng, D.
\newblock Numerical evaluation of the phase-field model for brittle fracture
  with emphasis on the length scale.
\newblock {\em Computational Mechanics} {\bf 2017}, {\em 59},~737--752.

\bibitem[Farrell and Maurini(2017)]{farrell2017linear}
Farrell, P.; Maurini, C.
\newblock Linear and nonlinear solvers for variational phase-field models of
  brittle fracture.
\newblock {\em International Journal for Numerical Methods in Engineering} {\bf
  2017}, {\em 109},~648--667.

\bibitem[Ambati \em{et~al.}(2015)Ambati, Gerasimov, and
  De~Lorenzis]{ambati2015review}
Ambati, M.; Gerasimov, T.; De~Lorenzis, L.
\newblock A review on phase-field models of brittle fracture and a new fast
  hybrid formulation.
\newblock {\em Computational Mechanics} {\bf 2015}, {\em 55},~383--405.

\bibitem[{CEN EN572-1}()]{en2004572}
CEN EN572-1: 2004 Glass in building-Basic soda lime silicate glass
  products-Part 1.
\newblock Standard, Deutsches Institut f{\"u}r Bautechnik, Berlin, Germany,
  2004.

\bibitem[Wiederhorn(1969)]{wiederhorn1969fracture}
Wiederhorn, S.M.
\newblock Fracture surface energy of glass.
\newblock {\em Journal of the American Ceramic Society} {\bf 1969}, {\em
  52},~99--105.

\bibitem[Wang \em{et~al.}(2017)Wang, Yang, Liu, Zhang, and
  Zhao]{wang2017comparative}
Wang, X.E.; Yang, J.; Liu, Q.F.; Zhang, Y.M.; Zhao, C.
\newblock A comparative study of numerical modelling techniques for the
  fracture of brittle materials with specific reference to glass.
\newblock {\em Engineering Structures} {\bf 2017}, {\em 152},~493--505.

\bibitem[Pham \em{et~al.}(2017)Pham, Ravi-Chandar, and
  Landis]{pham2017experimental}
Pham, K.; Ravi-Chandar, K.; Landis, C.
\newblock Experimental validation of a phase-field model for fracture.
\newblock {\em International Journal of Fracture} {\bf 2017}, {\em
  205},~83--101.

\bibitem[Wu and Nguyen(2018)]{wu2018length}
Wu, J.Y.; Nguyen, V.P.
\newblock A length scale insensitive phase-field damage model for brittle
  fracture.
\newblock {\em Journal of the Mechanics and Physics of Solids} {\bf 2018}, {\em
  119},~20--42.

\bibitem[Borden \em{et~al.}(2012)Borden, Verhoosel, Scott, Hughes, and
  Landis]{borden2012phase}
Borden, M.J.; Verhoosel, C.V.; Scott, M.A.; Hughes, T.J.; Landis, C.M.
\newblock A phase-field description of dynamic brittle fracture.
\newblock {\em Computer Methods in Applied Mechanics and Engineering} {\bf
  2012}, {\em 217},~77--95.

\bibitem[Logg \em{et~al.}(2012)Logg, Mardal, and Wells]{logg2012automated}
Logg, A.; Mardal, K.A.; Wells, G.
\newblock {\em Automated solution of differential equations by the finite
  element method: The FEniCS book}; Vol.~84, Springer Science \& Business
  Media,  2012.

\bibitem[Schmidt(2020)]{git_jarda}
Schmidt, J.
\newblock Laminated\_glass\_fracture\_QS supplementary code for {Phase-field}
  fracture modelling of thin monolithic or laminated glass plates under
  quasi-static bending.
\newblock \url{https://gitlab.com/JaraSit/laminated_glass_fracture_qs},  2020.

\bibitem[Bleyer \em{et~al.}(2017)Bleyer, Roux-Langlois, and
  Molinari]{bleyer2017dynamic}
Bleyer, J.; Roux-Langlois, C.; Molinari, J.F.
\newblock Dynamic crack propagation with a variational phase-field model:
  limiting speed, crack branching and velocity-toughening mechanisms.
\newblock {\em International Journal of Fracture} {\bf 2017}, {\em
  204},~79--100.

\bibitem[Natarajan \em{et~al.}(2019)Natarajan, Annabattula,
  Mart{\'\i}nez-Pa{\~n}eda, et~al.]{natarajan2019phase}
Natarajan, S.; Annabattula, R.K.; Mart{\'\i}nez-Pa{\~n}eda, E.; others.
\newblock Phase field modelling of crack propagation in functionally graded
  materials.
\newblock {\em Composites Part B: Engineering} {\bf 2019}, {\em 169},~239--248.

\bibitem[Hunt and Baker(1995)]{hunt1995principles}
Hunt, G.W.; Baker, G.
\newblock Principles of localization in the fracture of quasi-brittle
  structures.
\newblock {\em Journal of the Mechanics and Physics of Solids} {\bf 1995}, {\em
  43},~1127--1150.

\bibitem[Brocca(1997)]{brocca1997analysis}
Brocca, M.
\newblock Analysis of cracking localization and crack growth based on
  thermomechanical theory of localization.
\newblock PhD thesis, University of Tokyo, Tokyo,  1997.

\bibitem[Audy(2003)]{audy1993localization}
Audy, M.
\newblock Localization of inelastic deformation in problems freeof initial
  stress concentrators.
\newblock Master's thesis, Czech Technical University in Prague, Faculty of
  Civil Engineering, Prague,  2003.

\bibitem[{DIN 18008-1}()]{din2010glas}
DIN 18008-1:2010-12 Glass in Building - Design and construction rules - Part 1:
  Terms and general bases.
\newblock Standard, German Institute for Standardisation, Berlin, Germany,
  2010.

\bibitem[Duser \em{et~al.}(1999)Duser, Jagota, and Bennison]{Duser:1999:AGBL}
Duser, A.V.; Jagota, A.; Bennison, S.J.
\newblock Analysis of glass/polyvinyl butyral laminates subjected to uniform
  pressure.
\newblock {\em Journal of Engineering Mechanics} {\bf 1999}, {\em
  125},~435--442.
\newblock
  doi:{\changeurlcolor{black}\href{https://doi.org/10.1061/(ASCE)0733-9399(1999)125:4(435)}{\detokenize{10.1061/(ASCE)0733-9399(1999)125:4(435)}}}.

\bibitem[Christensen(2012)]{christensen2012theory}
Christensen, R.
\newblock {\em Theory of viscoelasticity: an introduction}; Elsevier,  2012.

\bibitem[H{\'a}na \em{et~al.}(2018)H{\'a}na, Eli{\'a}{\v{s}}ov{\'a}, and
  Sokol]{hanafour}
H{\'a}na, T.; Eli{\'a}{\v{s}}ov{\'a}, M.; Sokol, Z.
\newblock For point bending tests of double laminated glass panels.
\newblock  Proc. 24th International Conference Engineering Mechanics 2018,
  2018, Vol.~24, pp. 285--288.
\newblock
  doi:{\changeurlcolor{black}\href{https://doi.org/doi:10.21495/91-8-285}{\detokenize{doi:10.21495/91-8-285}}}.

\bibitem[Sargado \em{et~al.}(2018)Sargado, Keilegavlen, Berre, and
  Nordbotten]{sargado2018high}
Sargado, J.M.; Keilegavlen, E.; Berre, I.; Nordbotten, J.M.
\newblock High-accuracy phase-field models for brittle fracture based on a new
  family of degradation functions.
\newblock {\em Journal of the Mechanics and Physics of Solids} {\bf 2018}, {\em
  111},~458--489.

\end{thebibliography}
\end{document}